\documentclass[aps,prd,showpacs,preprintnumbers,amsmath,amssymb,nofootinbib]{revtex4}
\usepackage{graphicx}
\usepackage{verbatim}
\usepackage{slashed}
\usepackage{amsthm}
\usepackage{amsfonts}
\usepackage[dvips]{color}
\usepackage{axodraw}

\def\ththMat#1#2#3#4#5#6#7#8#9{\ensuremath{\begin{pmatrix}#1&#2&#3\\
                                                          #4&#5&#6\\
                                                          #7&#8&#9\end{pmatrix}}}
\def\twoneMat#1#2{\ensuremath{\begin{pmatrix}#1\\
                                              #2\end{pmatrix}}}
\def\twtwMat#1#2#3#4{\ensuremath{\begin{pmatrix}#1&#2\\
                                                #3&#4\end{pmatrix}}}
\def\onetwMat#1#2{\ensuremath{\begin{pmatrix}#1&#2\end{pmatrix}}}

\def\order#1{\ensuremath{\mathcal{O}\left(#1\right)}}

\def\BR#1{\ensuremath{\text{Br}(#1)}}
\def\drate#1{\ensuremath{\Gamma(#1)}}

\begin{document}

\title{Phenomenological constraints on minimally coupled exotic lepton triplets}
\author{Chun-Khiang~Chua}
\email{ckchua@cycu.edu.tw}
\affiliation{Department of Physics, Chung Yuan Christian University,
Chung-Li, Taiwan 320, Republic of China.}
\author{Sandy~S.~C.~Law}
\email{slaw@cycu.edu.tw}
\affiliation{Department of Physics, Chung Yuan Christian University,
Chung-Li, Taiwan 320, Republic of China.}

\date{14 February 2011}

\begin{abstract}
By introducing a set of new triplet leptons (with nonzero hypercharge) that can Yukawa couple to their Standard Model counterparts, new sources of tree-level flavor changing currents are induced via mixing. 
In this work, we study some of the consequences of such new contributions on processes such as the leptonic decays of gauge bosons, $\ell \rightarrow 3 \ell'$ and $\ell \rightarrow \ell' \gamma$ which violate lepton flavor, and $\mu$-$e$ conversion in atomic nuclei. Constraints are then placed on the parameters associated with the exotic triplets by invoking the current low-energy experimental data. Moreover, the new physics contribution to the lepton anomalous magnetic moments is calculated.
\end{abstract}

\maketitle


\section{Introduction}

The discovery of neutrino oscillations \cite{neutrinos_exp} has long been suggestive of new physics in the lepton sector. It provides compelling evidence for nonzero neutrino masses, and hints of possible lepton flavor violation (LFV). However, it is well-known that the minimal Standard Model (SM) cannot incorporate these new ingredients, so it must be extended in one way or the other as a result. Clearly, there is a huge variety of approaches for introducing new physics. Nevertheless, from the point of view of phenomenological studies, the most essential part of any models is the effective couplings induced between ordinary SM particles and the exotic ones. Therefore, even without specifying the underlying mechanisms (or UV completions) that give rise to these operators, a lot of useful analyses on the new particles can be studied. This is the approach we shall adopt in this work.

Whilst there are potentially many different new effective operators which can lead to interesting phenomenologies, our main focus here is motivated by the generic minimal couplings of the form
\begin{equation}\label{eqn:general_form}
 Y_\text{exotic} \;\text{(SM particle)}\cdot \text{(SM particle)} \cdot \text{(exotic particle)}\;,
\end{equation}
where $Y_\text{exotic}$ denotes the coupling strength. Such minimal interactions are of interest because it is relatively simple and may lead to well-defined collider signatures \cite{DelNobile:2009st} which may be seen at the LHC in the near future. 
Since we would like to concentrate on the lepton sector alone, we take all particles  in (\ref{eqn:general_form}) to be \emph{uncolored} (in the $SU(3)_c$ sense) but allowing the ``exotic particle'' to be either a scalar boson, a fermion or a vector boson. 
With these choices and the requirement of renormalizability, there are five distinct types of interaction with the SM fields (schematically)\footnote{We have ignored terms like (SM Higgs)(SM Higgs)(new boson) because they do not involve any type of leptons.},
\begin{eqnarray}\label{equ:5_types}
&\text{(i)}\;\;
 L_L \times L_L \times [\text{new}]\;,
 \;\;
\text{(ii)}\;\; 
 L_L \times \ell_R \times [\text{new}]\;,
  \;\;
\text{(iii)}\;\;  
 \ell_R \times \ell_R \times [\text{new}]\;, \nonumber \\
&\text{(iv)}\;\;  
 L_L \times \phi \times [\text{new}]\;,
 \;\;
\text{(v)}\;\; 
 \ell_R \times \phi \times [\text{new}] \;.
\end{eqnarray}
where $L_L =  (\nu_L, \ell_L)^T$ is the left-handed (LH) lepton doublet, $\ell_R$ is the right-handed (RH) lepton singlet, and $\phi=(\phi^+, \phi^0)^T$ denotes the SM Higgs doublet. Suppose these interaction terms must also obey Lorentz and SM gauge symmetries, then there are only 13 types of exotic multiplets (see Table~\ref{table:field_summary}) which fit either one of the setups in (\ref{equ:5_types}). 
Furthermore, it is perhaps obvious that the majority of the new particles implied by these minimal couplings
have already been closely studied due to other motivations. However, to the best of our knowledge, the exotic lepton triplets with \emph{nonzero} hypercharge, $E_{R,L}=(E^0_{L,R},E^{-}_{L,R},E^{--}_{L,R})^T$, and the doublets, $\tilde{L}_{L,R} = (\tilde{L}^-_{L,R},\tilde{L}^{--}_{L,R})^T$ (see Table~\ref{table:field_summary} for their transformation properties) have received very little attention.\footnote{Recently, the $E_{R,L}$-like triplets were mentioned in the context of neutrino mass generation involving a triply charged Higgs \cite{Babu:2009aq}.} 

So, the aim of this work is to fill part of that gap by investigating in some details the implications of introducing $E_{R,L}$ to the SM.\footnote{The analysis of the exotic doublets, $\tilde{L}_{L,R}$, shall be presented elsewhere.} We begin by elucidating the formalism used to analyze the system in the next section, 
before deriving various experimental constraints on the relevant new physics parameters in subsequent sections (with a summary of all constraints and fits collected in Sec.~\ref{sec:global_fit}). Processes such as $W$ and $Z$ decays (Sec.~\ref{sec:W_Z_decays}), LFV decays (e.g. $\ell\rightarrow 3\ell'$, $\ell\rightarrow \ell' \gamma$) and $\mu$-$e$ conversion in atomic nuclei (Sec.~\ref{sec:LFV_all}) are considered as a result, while discussion on the new physics contribution to the lepton anomalous magnetic moments will also be included (Sec.~\ref{sec:AMM}).

\begin{table}[ht]
\begin{center}
\begin{tabular}{|c|c|c|c|c|c|c|}
\hline
[new] & spin & $SU(2)_L$ & $U(1)_Y$ &  type & SM fields involved & studied in\\
\hline
$\Phi_i$ & 0 & 2 & $1/2$ & (ii) & $\overline{L}_L\;\; e_R$ & multi-Higgs doublet models \cite{multi_Higgs_models, rad_seesaw_eg}\\
$\chi_0$ & 0 & 1 & $-1$ &  (i) &$\overline{L}_L\;\; L_L^c$ & dilepton/Babu-Zee models \cite{rad_seesaw_eg, dileptons1, dileptons2, babu_zee}\\
$\Delta$ & 0 & 3 & $-1$ &  (i) &$\overline{L}_L\;\; L_L^c$ & dilepton/Type-II seesaw 
                                   \cite{dileptons1, doubly_Higg, type2seesaw, Abada:2007ux}\\
$\xi_0$ & 0 & 1 & 2 &  (iii) &$\overline{e}_R\;\; e_R^c$ & dilepton/Babu-Zee models \cite{dileptons1, dileptons2, babu_zee, babu_zee2}
\\%
\hline
$\nu_R$ & 1/2 & 1 & $0$ &  (iv) &$\overline{L}_L\;\; \phi^c$ & Type-I seesaw \cite{Abada:2007ux, type1seesaw, Biggio:2008in, type1_3, type1set2}\\
$\Sigma_R$ & 1/2 & 3 & $0$ &  (iv) &$\overline{L}_L\;\; \phi^c$ & Type-III seesaw \cite{Abada:2007ux, type1_3, type3seesaw, Abada:2008ea}\\
$L_L''$ & 1/2 & 2 & $-1/2$ &  (v) &$\overline{e}_R\;\; \phi^\dagger$ & 4th generation leptons \cite{4thgen}\\
$\ell_R''$ & 1/2 & 1 & $-1$ & (iv) &$\overline{L}_L\;\; \phi$ & 4th generation leptons \cite{4thgen}\\
$E_{R,L}$ & 1/2 & 3 & $-1$ & (iv) &$\overline{L}_L\, \phi\;$  ($E_R$ only) & --- \emph{rarely discussed} ---\\
$\tilde{L}_{L,R}$ & 1/2 & 2 & $-3/2$ &  (v) & $\overline{e}_R \,(\phi^{c})^{\dagger}$ ($\tilde{L}_L$ only)& --- \emph{rarely discussed} ---\\
\hline
$Z_{\mu}'$ & 1 & 1 & $0$ & (i) \& (iii) & $\overline{L}_L\;\; L_L\;$ \& $\;\overline{e}_R\;\; e_R$& \\
$X_{\mu}$ & 1 & 2 & $-3/2$ & (ii) & $\overline{L}_L\;\; e_R^c$& GUT/dilepton boson models \cite{dileptons1, dilepton_boson}\\
$W_{\mu}'$ & 1 & 3 & $0$ & (i) &$\overline{L}_L\;\; L_L$ & \\
\hline
\end{tabular}
\caption{
Summary of the 13 types of exotic multiplets induced by the 5 general types of minimal couplings: 
(i) $L_L \times L_L \times [\text{new}]$, 
(ii) $L_L \times \ell_R \times [\text{new}]$,
(iii) $\ell_R \times \ell_R \times [\text{new}]$,
(iv) $L_L \times \phi \times [\text{new}]$,
and
(v) $\ell_R \times \phi \times [\text{new}]$.
Hypercharges are defined with $Q=I_3+Y$.}
\label{table:field_summary}
\end{center}
\end{table}

\section{Model with exotic lepton triplets, $E_{R,L}$}\label{sec:the_model}

In order to identify and study the new phenomenologies arising from the mixing with the exotic lepton triplets (and to establish the notations), we shall begin by describing the model in detail. Consider adding to the SM two sets  of new leptons (RH plus LH) which transform as triplets in $SU(2)_L$ and all carrying hypercharge of $-1$ (where we have defined $Q=I_3+Y$). We can conveniently group them together in a $2\times 2$ matrix representation as follows
\begin{equation}
 E_R =\twtwMat{E_R^- / \sqrt{2}}{E_R^0}
              {E_R^{--}}{-E_R^- / \sqrt{2}}\;,
 \qquad            
 E_L =\twtwMat{E_L^- / \sqrt{2}}{E_L^0}
              {E_L^{--}}{-E_L^- / \sqrt{2}}\;, 
\end{equation}
%
%
%
where $E_R$ and $E_L$ are independent fields and both transform as $(1,3,-1)$ under the SM gauge group.\footnote{As a result of the identical transformation properties for RH and LH fields, chiral anomalies cancel automatically.} In the following, we shall also introduce three RH neutrino fields, $\nu_R$, so that neutrinos can have a Dirac mass. However, we do \emph{not} include a Majorana mass term (e.g. $\overline{\nu_R^c} M_R \nu_R$) in the Lagrangian (and hence no seesaw mechanism\footnote{A full discussion on the mixing effects due to seesaw models can be found in \cite{Abada:2007ux, Antusch:2006vwa}.}) for simplicity. Thus, the Lagrangian of interest is given by
\begin{equation}\label{equ:main_L}
 \mathcal{L}^E = 
  \text{Tr} \left[\overline{E}_R i\slashed{D} E_R\right]+
  \text{Tr} \left[\overline{E}_L i\slashed{D} E_L\right]-
  \text{Tr} \left[\overline{E}_R M_E E_L + h.c.\right]
  - \left[
   \overline{L}_L Y_E E_R \phi
  + \overline{L}_L Y_\ell \phi \,\ell_R
  + \overline{L}_L Y_\nu \phi^c \,\nu_R
  + h.c.
  \right]\;,
\end{equation}
where $\phi^c = (\phi^{0*}, -\phi^-)^T$, and the covariant derivative,
\begin{equation}
\slashed{D} = \slashed{\partial} 
               -\frac{ig}{\sqrt{2}} \left[\slashed{W}^+ \twtwMat{0}{1}{0}{0}
               +
               \slashed{W}^- \twtwMat{0}{0}{1}{0}
               \right]
               -\frac{ig}{\cos \theta_w} \slashed{Z} (I_3 - \sin^2 \theta_w Q)
               +i e \slashed{A} Q\;, \quad (e>0)\;,
\end{equation}
with $Q$ and $I_3$ being the operators for electric charge and the 3rd component of isospin respectively.
In (\ref{equ:main_L}), the Yukawa term involving $Y_E$ defines the minimal coupling between SM leptons and $E_R$ while $M_E$ sets the energy scale of the new physics. It is worth pointing out that SM symmetries forbid a similar type of minimal couplings for $E_L$ with other SM leptons, and hence $E_L$ enters into this picture only via the mass terms.
Writing out all the relevant interactions in (\ref{equ:main_L}), we have
\begin{equation}
 \mathcal{L}^E = 
 \mathcal{L}^W+
 \mathcal{L}^Z+
 \mathcal{L}^\text{mass}+
 \mathcal{L}^H 
 + \cdots
 \;,
\end{equation}
where
\begin{align}
 \mathcal{L}^W &=
  \frac{g}{\sqrt{2}} \left[
   \overline{\nu}_L \slashed{W}^+ \ell_L
    + \frac{1}{\sqrt{2}}
    \left[ 
     \overline{E_R^-} \slashed{W}^+ E_R^{--}
     -\overline{E_R^0} \slashed{W}^+ E_R^{-}
     +\{E_R \rightarrow E_L\}
     \right]
   \right]
   + h.c.\;, \label{equ:L_W}\\
  \mathcal{L}^Z &=
  \frac{g}{\cos \theta_w} \left[
   \frac{1}{2}\; \overline{\nu}_L \slashed{Z} \nu_L
   +\left(-\frac{1}{2} +  \sin^2 \theta_w\right) \overline{\ell}_L \slashed{Z} \ell_L
   +\sin^2 \theta_w \;\overline{\ell}_R \slashed{Z} \ell_R
   \right.\nonumber\\
   &\qquad\qquad\quad\left.
   + \sin^2 \theta_w \;\overline{E_R^-} \slashed{Z} E_R^-
   + \left(-1+\sin^2 \theta_w\right) \;\overline{E_R^{--}} \slashed{Z} E_R^{--}
   + \overline{E_R^0} \slashed{Z} E_R^0
   +\{E_R\rightarrow E_L\}
   \right]\;,\label{equ:L_Z}\\
  \mathcal{L}^\text{mass} &=   
  -\overline{E_R^-} M_E E_L^{-}
  -\overline{E_R^0} M_E E_L^{0}
  -\overline{E_R^{--}} M_E E_L^{--}
  -\frac{v}{\sqrt{2}} \overline{\nu}_L Y_E E_R^{0}
  +\frac{v}{2} \overline{\ell}_L Y_E E_R^{-}
  -\overline{\ell}_L m_\ell \ell_R
  -\overline{\nu}_L m_D \nu_R + h.c.\;, \label{equ:L_mass}\\
  \mathcal{L}^H &=
  -\frac{1}{\sqrt{2}}\overline{\nu}_L Y_E E_R^0 H
  +\frac{1}{2}\overline{\ell}_L Y_E E_R^- H
  -\frac{1}{v}\overline{\ell}_L m_\ell \ell_R H
  -\frac{1}{v}\overline{\nu}_L m_D \nu_R H + h.c.\;. \label{equ:L_H}
\end{align}
In getting (\ref{equ:L_mass}) and (\ref{equ:L_H}), we have written $\phi=(\phi^+, \phi^0)^T \equiv (\phi^+, (v + H + i\eta)/\sqrt{2})^T$, where $v$ is the Higgs vacuum expectation value, $\eta$ and $\phi^{\pm}$ are the would-be Goldstone bosons. Also, we have defined $m_\ell \equiv v Y_\ell/\sqrt{2}$ and $m_D\equiv v Y_\nu/\sqrt{2}$.

To deduce the mixing between SM leptons and the components of the exotic triplet, it is convenient to package the LH and RH  fields in the following way
\begin{equation}
 \twoneMat{\nu_L}{E_L^0} \;,\quad
 \twoneMat{\nu_R}{E_R^0} \;,\quad
 \twoneMat{\ell_L}{E_L^-} \;,\quad
 \twoneMat{\ell_R}{E_R^-}\;,
\end{equation}
and rewriting (\ref{equ:L_W}) to (\ref{equ:L_H}) in matrix forms. In particular, for $\mathcal{L}^\text{mass}$ we obtain
\begin{equation}
\mathcal{L}^\text{mass} =
 -\onetwMat{\overline{\nu}_R}{\overline{E_R^0}}
  \twtwMat{m_D^\dagger}{0}{v Y_E^\dagger/\sqrt{2}}{M_E}
   \twoneMat{\nu_L}{E_L^0}
 -\onetwMat{\overline{\ell}_R}{\overline{E_R^-}}
  \twtwMat{m_\ell}{0}{-v Y_E^\dagger/2}{M_E}
   \twoneMat{\ell_L}{E_L^-} 
   +h.c.\;. \label{equ:L_mass_matrix}
\end{equation}
Without loss of generality, one can choose to work in the basis where $m_\ell$ and $M_E$ are real and diagonal (which is what we have already assumed in writing out (\ref{equ:L_mass_matrix}) above). All fields are related to their mass eigenbasis via the unitary transformations
\begin{equation} \label{equ:UV_trans}
 \twoneMat{\ell_{L,R}}{E_{L,R}^-}
 = U_{L,R}
 \twoneMat{\ell_{L,R}}{E_{L,R}^-}_m\;,
 \qquad
 \twoneMat{\nu_{L,R}}{E_{L,R}^0} 
 = V_{L,R}
 \twoneMat{\nu_{L,R}}{E_{L,R}^0}_m \;,
\end{equation}
where the subscript $m$ indicates the mass basis. In general, $U_{L,R}$ and $V_{L,R}$ are $(3+n)\times (3+n)$ matrices with $n$ denoting the number of generations for the exotic $E_{L,R}$ fields. To $\order{v^2 M_E^{-2}}$, the transformation matrices are given by\footnote{In the definition of $V_R$, the neutrino right diagonalisation matrix $U_{\nu R}$ (from $U_\nu^\dagger m_D' U_{\nu R} \equiv m_D^\text{diag}$) has already been absorbed into $m_D$ in (\ref{equ:L_mass_matrix}). In other words, $m_D \equiv m_D' U_{\nu R}$.} 
\begin{align}
 U_L = \twtwMat{1-\lambda}{-v Y_E M_E^{-1}/2}{v M_E^{-1} Y_E^\dagger /2}{1-2\lambda'} \;,
 &\qquad
 U_R = \twtwMat{1}{-v\, m_\ell Y_E M_E^{-2}/2}{v M_E^{-2} Y_E^\dagger m_\ell /2}{1} \;,\label{equ:U_trans}\\
 V_L = \twtwMat{(1-2\lambda)U_\nu}{v Y_E M_E^{-1}/\sqrt{2}}
             {-v M_E^{-1} Y_E^\dagger U_\nu /\sqrt{2}}{1-4\lambda'} \;,
 &\qquad
 V_R = \twtwMat{1}{v\, m_D^\dagger Y_E M_E^{-2}/\sqrt{2}}{-v M_E^{-2} Y_E^\dagger m_D /\sqrt{2}}{1} \;,\label{equ:V_trans}
\end{align}
where
\begin{equation}\label{eqn:def_lambda}
 \lambda \equiv \frac{v^2}{8} \,Y_E\, \frac{1}{M_E^2}\, Y_E^\dagger \;,
 \quad \text{ and }\quad
 \lambda' \equiv \frac{v^2}{8} \frac{1}{M_E}\, Y_E^\dagger\, Y_E \frac{1}{M_E}\,\;, 
\end{equation}
are $3\times 3$ and $n\times n$ matrices in flavor space respectively, while $U_\nu$ is the unitary matrix that transforms $\nu_L$ into its mass eigenbasis. At this order, $U_\nu$ may be identified as the usual neutrino mixing matrix, $U_\text{PMNS}$.

Hence, $\mathcal{L}^W, \mathcal{L}^Z$ and $\mathcal{L}^H$ with respect to the mass eigenbasis become
\begin{align}
 \mathcal{L}^W &= \frac{g}{\sqrt{2}} \left\{
  \onetwMat{\overline{\nu}}{\overline{E^0}}_m \slashed{W}^+
   \left[P_L \,g_L^{CC} +P_R \,g_R^{CC}\right] 
   \twoneMat{\ell}{E^-}_m 
   +
  \onetwMat{\overline{\ell}}{\overline{E^-}}_m \slashed{W}^-
   \left[P_L (g_L^{CC})^\dagger +P_R (g_R^{CC})^\dagger\right] 
   \twoneMat{\nu}{E^0}_m 
   \right\}
    , \label{equ:L_CC} \\
  \mathcal{L}^Z &= \frac{g}{\cos \theta_w}
  \left[
     \onetwMat{\overline{\ell}}{\overline{E^-}}_m \slashed{Z}
   \left[P_L \,g_{L\ell}^{NC} +P_R \,g_{R\ell}^{NC}\right] 
   \twoneMat{\ell}{E^-}_m
   +
     \onetwMat{\overline{\nu}}{\overline{E^0}}_m \slashed{Z}
   \left[P_L \,g_{L\nu}^{NC} +P_R \,g_{R\nu}^{NC}\right] 
   \twoneMat{\nu}{E^0}_m 
   \right]
   \;,  \label{equ:L_NC}\\
   \mathcal{L}^H &= 
     \onetwMat{\overline{\ell}}{\overline{E^-}}_m H
   \left[P_L \left(g_{\ell}^{H}\right)^\dagger + P_R \,g_{\ell}^{H}\right] 
   \twoneMat{\ell}{E^-}_m
   +
     \onetwMat{\overline{\nu}}{\overline{E^0}}_m H
   \left[P_L \left(g_{\nu}^{H}\right)^\dagger + P_R \,g_{\nu}^{H}\right] 
   \twoneMat{\nu}{E^0}_m 
   \label{equ:L_higgs}\;,
\end{align}
with the new generalized coupling matrices given by (to leading order)
\begin{align}
 g_{L}^{CC} &= 
 \twtwMat{U_\nu^\dagger (1-\lambda)}{0}
   {vM_E^{-1} Y_E^\dagger /2\sqrt{2}}{(2\lambda' -1)/\sqrt{2}} \;, \qquad\qquad
 g_{R}^{CC} = \twtwMat{0}{v\,m_D^\dagger Y_E M_E^{-2}/2}
   {-v M_E^{-2} Y_E^\dagger m_\ell/2\sqrt{2}}{-1/\sqrt{2}} \;, 
 \label{equ:g_LCC}\\
 g_{L\ell}^{NC} &= \twtwMat{-1/2 + \sin^2 \theta_w + \lambda}{v Y_E M_E^{-1}/4}
 {v M_E^{-1} Y_E^\dagger/4}{-\lambda' + (1-2\lambda')\sin^2 \theta_w} \;, \quad
 g_{R\ell}^{NC} = \twtwMat{\sin^2 \theta_w}{0}{0}{\sin^2 \theta_w} \;, \label{equ:g_le_NC}\\
 g_{L\nu}^{NC} &= \twtwMat{1/2 + 2 U_\nu^\dagger \lambda U_\nu}
    {-v\, U_\nu^\dagger Y_E M_E^{-1}/2\sqrt{2}}{-v M_E^{-1} Y_E^\dagger U_\nu/2\sqrt{2}}
    {1-6\lambda'} \;,\quad
 g_{R\nu}^{NC} = \twtwMat{0}{-v\, m_D^\dagger Y_E M_E^{-2}/\sqrt{2}}
      {-v M_E^{-2} Y_E^\dagger m_D /\sqrt{2}}{1} \;, \label{equ:g_nu_NC}\\
 g_{\ell}^{H} &= \twtwMat{(3\lambda -1) m_\ell/v}{(1-\lambda)Y_E/2 + m_\ell^2 Y_E M_E^{-2}/2}
    {M_E^{-1} Y_E^\dagger m_\ell /2}{-v M_E^{-1} Y_E^\dagger Y_E/4} \;, \nonumber\\
 g_{\nu}^{H} &= \twtwMat{U_\nu^\dagger(6\lambda -1) m_D/v}
  {U_\nu^\dagger \left( -m_D m_D^\dagger Y_E M_E^{-2} + (1-2\lambda ) Y_E\right) /\sqrt{2}}
  {-M_E^{-1} Y_E^\dagger m_D/ \sqrt{2}}{-v M_E^{-1} Y_E^\dagger Y_E/2} \;.
  \label{equ:g_H_both}
\end{align}
Note that each upper-left $(3\times 3)$-block in (\ref{equ:g_LCC}) to (\ref{equ:g_H_both}) corresponds to the modified mixing matrix for the respective interaction involving SM leptons. In particular, we observe that new contributions to tree-level flavor changing currents would be provided by the nonzero off-diagonal entries of matrix $\lambda$. Furthermore, these $(3\times 3)$-submatrices that define the new mixings between ordinary leptons are now in general non-unitary. 

Suppose we define the non-unitary mixing matrix which is responsible for charged current mixing as
\begin{equation}\label{eqn:N_defn}
 N \equiv (1-\lambda)U_\nu \;,
\end{equation}
then we note that, at first order in $\lambda$, observable effects mediated by $W$ and $Z$ may be conveniently re-cast as follows
\begin{align}
  \mathcal{L}^{CC} &= \frac{g}{\sqrt{2}}\, \overline{\nu}\, \slashed{W}^+ P_L\, N^\dagger\, \ell +h.c.\;,
 \label{equ:L_CC_with_N}
  \\
  \mathcal{L}^{NC} &= \frac{g}{\cos \theta_w}\left\{ \overline{\ell} \, \slashed{Z}\, 
  \left[P_L 
  \left(-\frac{1}{2}\, N N^\dagger + \sin^2\theta_w \right) 
  +P_R \sin^2\theta_w
  \right]\ell
  + 
  \overline{\nu} \, \slashed{Z}\, P_L 
  \left(\frac{1}{2}\, (N^\dagger N)^{-2} \right) \nu
  \right\}\;.
   \label{equ:L_NC_with_N}
\end{align}
Expressions (\ref{equ:L_CC_with_N}) and (\ref{equ:L_NC_with_N}) are analogous to those derived and subsequently analyzed in \cite{Abada:2007ux} for seesaw models. It is worth pointing out that the structure displayed in (\ref{equ:L_NC_with_N}), though looks similar to its counterpart in  \cite{Abada:2007ux}, they are definitely \emph{not} identical in form. The small difference comes from the fact that these exotic triplets carry nonzero hypercharges which resulted in the $I_3$ assignments being different from the seesaw situations.

Although this viewpoint of linking the new physics to the non-unitary in weak mixing can be useful, for the purpose of our investigation here, we have found it more convenient to use the expressions written in (\ref{equ:L_CC}) to (\ref{equ:L_higgs}) for doing the calculations. Henceforth, we shall present all our discussions in terms of $\lambda$ rather than $N$.

\section{Constraints from $W$ and $Z$ decays}\label{sec:W_Z_decays}

As hinted earlier, elements of the $\lambda$ matrix (see (\ref{eqn:def_lambda}) for definition) which encapsulate all the essential information regarding triplets $E$, are the key to any new physics contributions to the electroweak processes considered in this paper. Amongst them, the most basic interactions are the tree-level $W$ and $Z$ decays into SM leptons. 
As we shall see, these processes can provide constraints on all elements of $\lambda$ although the restriction for the off-diagonal entries are not as stringent as those obtained from other LFV interactions (see Sec.~\ref{sec:LFV_all}). Nonetheless, their constraints for the diagonal elements of $\lambda$ will be useful in the later analysis of the anomalous magnetic moments (see Sec.~\ref{sec:AMM}).

\subsection{$W$ decays}\label{subsec:W}

The rate for $W$ decaying into a lepton of flavor $\alpha$ plus a neutrino may be straightforwardly obtained by invoking the relevant interaction terms in (\ref{equ:L_CC}). Using the usual approximation of negligible final lepton masses and in the centre of mass frame, one gets
\begin{align}
 \Gamma(W\rightarrow \ell_\alpha \nu_\alpha) &\equiv 
    \sum_i \Gamma(W\rightarrow \ell_\alpha \nu_i)\;, \nonumber \\
 &\simeq \frac{G_F M_W^3}{6\sqrt{2}\,\pi} (1- 2\lambda_{\alpha\alpha})\;, \label{eqn:W_decay_basic}
\end{align}
where we have only kept the leading order terms in $\lambda$. In (\ref{eqn:W_decay_basic}), $M_W$ is the mass of $W$ while $G_F$ is the Fermi constant extracted from muon decay when assuming \emph{only} SM physics.

In order for (\ref{eqn:W_decay_basic}) to be a useful bound on the elements of $\lambda$, one must also study the modification to the value of ``$G_F$'' as measured from muon decay experiments 
($\mu \rightarrow e$ + missing energy) in the presence of the new physics due to triplets $E$. It is not difficult to see from (\ref{equ:g_LCC}) and (\ref{equ:g_le_NC}) that additional tree-level flavor changing currents mediated by $W$ and $Z$ are expected to give rise to a new definition for the Fermi constant. In terms of the SM version of $G_F$, we have to leading order
\begin{equation} \label{eqn:modified_GF}
 G_F' \simeq G_F \sqrt{1-2 \lambda_{ee} -2\lambda_{\mu\mu}}\;, \quad\text{with}\quad G_F^2 \equiv g^4/ 32M_W^4\;.
\end{equation}
Using (\ref{eqn:modified_GF}) in (\ref{eqn:W_decay_basic}), one can rearrange the expression to obtain a global constraint on $\lambda_{ee}, \lambda_{\mu\mu}$ and $\lambda_{\tau\tau}$ in terms of experimental parameters:
\begin{align}
 K_{\alpha}\equiv
 \frac{(1-2\lambda_{\alpha\alpha})}{\sqrt{1-2 \lambda_{ee} -2\lambda_{\mu\mu}}} \simeq 
 \frac{\Gamma(W\rightarrow \ell_\alpha \nu_\alpha)\,6\sqrt{2} \,\pi}{G_F' M_W^3}\;,
 \qquad \alpha = e,\mu, \tau\;.
 \label{eqn:Ka_defn}
\end{align}
Putting in the respective values from \cite{Nakamura:2010zzi},\footnote{Note that what we have labelled as $G_F'$ is simply the experimentally measured Fermi constant $\approx 1.16637(1) \times 10^{-5}$ GeV$^{-2}$.} we arrive at the following bounds for $K_{\alpha}$:
\begin{equation}\label{eqn:Ka_values}
K_\alpha =
 \begin{cases}
  0.986\pm 0.033\;,\quad \alpha = e\;, \\
  0.969\pm 0.034\;,\quad \alpha = \mu\;, \\
  1.032\pm 0.040\;,\quad \alpha = \tau\;.
 \end{cases}
\end{equation}
As expected, this quantity (within experimental uncertainties) is very close to 1 --- the limit where the new physics is decoupled.

\subsection{$Z$ decays}\label{subsec:Z}

In this subsection, we investigate the bounds for the elements of $\lambda$ coming from $Z$ decaying into charged leptons: $Z \rightarrow \ell_\alpha \overline{\ell}_\beta$. The $\alpha=\beta$ cases will place restrictions on $\lambda_{\alpha\alpha}$'s whereas for $\alpha \neq \beta$, the off-diagonal entries can be constrained.

Applying the usual formalism on the modified couplings in (\ref{equ:g_le_NC}), the decay rate, $\Gamma(Z\rightarrow \ell_\alpha \overline{\ell}_\alpha)$, may be easily written down (in the usual massless limit for the final state leptons) as
\begin{align}
 \Gamma(Z\rightarrow \ell_\alpha \overline{\ell}_\alpha) &=
  \frac{G_F M_Z^3}{3 \sqrt{2} \,\pi}\left(
  \left|(g_{L\ell}^{NC})^{11}_{\alpha\alpha}\right|^2  +\left|(g_{R\ell}^{NC})^{11}_{\alpha\alpha}\right|^2
  \right)\;, \nonumber\\
  &\simeq
  \frac{G_F' M_Z^3}{3\pi \sqrt{2-4\lambda_{ee}-4\lambda_{\mu\mu}}}\left(
  \left|\sin^2 \theta_w -\frac{1}{2} +\lambda_{\alpha\alpha}\right|^2  +\left|\sin^2 \theta_w\right|^2
  \right), \quad \alpha = e, \mu, \tau\;, \label{eqn:Z_aa_rate}
\end{align}
where we have again included the correction to the Fermi constant. $\theta_w$ and $M_Z$ are the usual Weinberg angle and $Z$ boson mass respectively. Putting the decay widths obtained from experiments \cite{Nakamura:2010zzi} into (\ref{eqn:Z_aa_rate}) for each lepton flavor $\alpha$, one gets a system of three equations in the $\lambda_{\alpha\alpha}$'s. Solving these simultaneously then yields
\begin{align}
\lambda_{\alpha\alpha} =
 \begin{cases}
  -2.7\mp 0.4\times 10^{-3}\;,\quad \alpha = e\;, \\
  -2.9\mp 0.4\times 10^{-3}\;,\quad \alpha = \mu\;, \\
  -3.1\mp 0.4\times 10^{-3}\;,\quad \alpha = \tau\;.
 \end{cases} \label{eqn:Laa_values}
\end{align}
These results should be checked against the values obtained in the $W$ decays for consistency. Taking into account the uncertainties in $K_\alpha$, we have found that the bounds displayed in (\ref{eqn:Laa_values}) are compatible with those in (\ref{eqn:Ka_values}). Although one may worry about the negative sign in front of $\lambda_{\alpha\alpha}$, this outcome is not unexpected given that there is also a minus sign in the definition of (\ref{eqn:N_defn}). 

Next, we turn our attention to the case where $\alpha \neq \beta$. The decay rate is given by
\begin{align}\label{eqn:Zab_rate}
 \Gamma(Z\rightarrow \ell_\alpha \overline{\ell}_\beta) &=
 \frac{G_F M_Z^3}{3 \sqrt{2} \,\pi} \left|\lambda_{\alpha\beta} \right|^2\;, \qquad \alpha \neq \beta \;.
\end{align}
Clearly, in the limit $\lambda_{\alpha\beta} \rightarrow 0$, this rate disappears. This is in accordance with the fact that there is no flavor changing neutral currents (FCNC) at tree-level in the SM.
Writing this as a branching ratio and keeping only the leading order terms in the denominator, one has
\begin{align}
 \BR{Z\rightarrow \ell_\alpha \overline{\ell}_\beta} &=
  \frac{\drate{Z\rightarrow \ell_\alpha \overline{\ell}_\beta}}
  {\Gamma(Z\rightarrow \ell_\sigma \overline{\ell}_\sigma)}\,\BR{Z\rightarrow \ell_\sigma \overline{\ell}_\sigma}\;,
    \nonumber
 \\
 &\simeq \frac{\left|\lambda_{\alpha\beta} \right|^2\BR{Z\rightarrow \ell_\sigma \overline{\ell}_\sigma}}{2 \sin^4 \theta_w - \sin^2 \theta_w + 1/4}
    \;.\label{eqn:Zab_BR}
\end{align}
From this, we can derive the following bounds for $|\lambda_{\alpha\beta}|$:\footnote{Note that the LFV branching ratios quoted in \cite{Nakamura:2010zzi} is in fact the experimental values for $\BR{Z\rightarrow \ell_\alpha \overline{\ell}_\beta} + \BR{Z\rightarrow \overline{\ell}_\alpha \ell_\beta }$. Therefore, the expression in (\ref{eqn:Zab_BR}) must be multiplied by a factor of 2 before applying the experimental numbers.}
\begin{align}
  &|\lambda_{e\mu}|< 1.8 \times 10^{-3}\;, \label{eqn:Z_offd_1}\\
 &|\lambda_{e\tau}|< 4.3 \times 10^{-3}\;,\label{eqn:Z_offd_2}\\
&|\lambda_{\mu\tau}|< 4.7 \times 10^{-3}\;.\label{eqn:Z_offd_3}
\end{align}
Notice that since $\lambda$ is hermitian (as we are working in the basis where $M_E$ is real and diagonal),  $|\lambda_{\alpha\beta}|=|\lambda_{\beta\alpha}|$ necessarily holds.

\section{Constraints from LFV decays of charged leptons and $\mu$-$e$ conversion in atomic nuclei}\label{sec:LFV_all}

Some of the strongest constraints on the new physics come from the studies of lepton flavor violating decays of ordinary charged leptons. Therefore, in the following two subsections, we present our analysis of the contributions induced by the exotic triplets on charged lepton processes like $\ell \rightarrow 3\ell'$ and $\ell \rightarrow \ell' \gamma$. Furthermore, in the third subsection, we shall take a look at the bound coming from experiments studying the muon-to-electron conversion in atomic nuclei as it is well-known \cite{Bernabeu:1993ta} that such process can give rise to a very strong constraint on the $\mu$-$e$-$Z$ vertex.

\subsection{tree-level $\ell \rightarrow 3\ell'$ decays}\label{subsec:l_3l}

Given three generations of ordinary leptons, there are only three generic types of final lepton states possible for a charged lepton decaying into three lighter ones: $\ell_\beta \ell_\beta \overline{\ell}_\beta$, 
$\ell_\sigma \ell_\beta \overline{\ell}_\beta$ and 
$\overline{\ell}_\sigma \ell_\beta \ell_\beta$, where $\beta\neq\sigma\neq\alpha$, with $\alpha$ denoting the flavor of the decaying lepton. For all of these cases, the mediating particle can be either the gauge  boson $Z$ or the Higgs boson $H$. However, the amplitude associated with the Higgs is suppressed by a factor of $m_\alpha^2/M_H^2$, where $m_\alpha$ and $M_H$ denote the lepton and Higgs masses respectively. Thus, we may ignore their contributions to a good approximation.

Extracting the relevant coupling from (\ref{equ:g_le_NC}), and invoking the usual assumption of negligible final state masses, we get the following formulae for the branching ratios:
\begin{align}
 \BR{\ell_\alpha \rightarrow \ell_\beta \ell_\beta \overline{\ell}_\beta} 
    &= \frac{\drate{\ell_\alpha \rightarrow \ell_\beta \ell_\beta \overline{\ell}_\beta}}
    {\drate{\ell_\alpha \rightarrow \ell_\beta \nu_\alpha \overline{\nu}_\beta}}\,
    \BR{\ell_\alpha \rightarrow \ell_\beta \nu_\alpha \overline{\nu}_\beta}\;, \nonumber\\
    &\simeq |\lambda_{\beta\alpha}|^2 (12\sin^4\theta_w -8\sin^2\theta_w +2)\,
    \BR{\ell_\alpha \rightarrow \ell_\beta \nu_\alpha \overline{\nu}_\beta}\;,
    &\text{ for } \alpha = \mu, \tau\;, 
    \label{eqn:BR_abbb}\\
\intertext{and}
   \BR{\ell_\alpha \rightarrow \ell_\sigma \ell_\beta \overline{\ell}_\beta} 
    &\simeq |\lambda_{\sigma\alpha}|^2 (8\sin^4\theta_w -4\sin^2\theta_w +1)\,
    \BR{\ell_\alpha \rightarrow \ell_\beta \nu_\alpha \overline{\nu}_\beta}\;,&
    \label{eqn:BR_asbb} 
   \\
   \BR{\ell_\alpha \rightarrow \overline{\ell}_\sigma \ell_\beta \ell_\beta}
    &\simeq 2 |\lambda_{\beta\sigma}|^2 |\lambda_{\beta\alpha}|^2 \,
    \BR{\ell_\alpha \rightarrow \ell_\beta \nu_\alpha \overline{\nu}_\beta}\;,
    &\text{ for } \alpha = \tau \text{ only}\;, 
    \label{eqn:BR_absb}
\end{align}
where we have kept only the leading order terms.

For (\ref{eqn:BR_abbb}), there are three kinematically allowed processes 
($\mu \rightarrow 3e, \tau \rightarrow 3e, \tau \rightarrow 3\mu$), which lead to the constraints
\begin{align}
 &|\lambda_{e\mu}|< 1.1 \times 10^{-6}\;, \label{eqn:Lee_bound_tree}\\
 &|\lambda_{e\tau}|< 5.0 \times 10^{-4}\;,\label{eqn:Let_bound_tree}\\
&|\lambda_{\mu\tau}|< 4.8 \times 10^{-4}\;,\label{eqn:Lmt_bound_tree}
\end{align}
while (\ref{eqn:BR_asbb}) has two possibilities ($\tau \rightarrow e \mu\overline{\mu}$ and 
$\tau \rightarrow \mu e\overline{e}$), yielding
\begin{align}
 &|\lambda_{e\tau}|<6.5 \times 10^{-4} \;,\\
&|\lambda_{\mu\tau}|< 5.5 \times 10^{-4}\;.
\end{align}
Finally, we have 
\begin{align}
 &|\lambda_{\mu e}||\lambda_{\mu\tau}|< 2.6 \times 10^{-4}\;,\\
&|\lambda_{e\mu}||\lambda_{e\tau}|< 2.4 \times 10^{-4}\;,\label{eqn:LemLet_bound_tree}
\end{align}
from another two possibilities ($\tau \rightarrow \overline{e}\mu\mu$ and  
$\tau \rightarrow \overline{\mu}e e$) allowed by (\ref{eqn:BR_absb}). Note that in deriving (\ref{eqn:Lee_bound_tree}) to (\ref{eqn:LemLet_bound_tree}), we have used the branching ratios from \cite{Nakamura:2010zzi}.

As expected, these LFV processes provide a stronger set of constraints than those derived in (\ref{eqn:Z_offd_1}) to (\ref{eqn:Z_offd_3}) from the previous section.

\subsection{radiative $\ell \rightarrow \ell'\gamma$ decays via one loop}\label{subsec:l_lg}

Another type of LFV processes that has received enormous amount of attention is the radiative decays of charged leptons ($\ell \rightarrow \ell'\gamma$). There is continually much experimental effort  on improving the bounds associated with these rare interactions.\footnote{For a review, see for example \cite{Marciano:2008zz}.} The current MEG experiment \cite{exp_MEG} located at PSI is expected to reach a sensitivity of \order{10^{-13}} for the $\mu \rightarrow e \gamma$ branching ratio, which is a significant improvement compare to the current limit of \BR{\mu\rightarrow e\gamma} $< 1.2 \times 10^{-11}$ \cite{Brooks:1999pu}. In addition, the Super KEKB project \cite{exp_SuperB} will provide the platform for investigating LFV $\tau$ decays at an unprecedented precision. As a result, the bounds on $\tau \rightarrow e\gamma$ and $\tau\rightarrow \mu\gamma$ are also expected to tighten.

\begin{figure}[tb]
\begin{center}
\includegraphics[width=0.25\columnwidth]{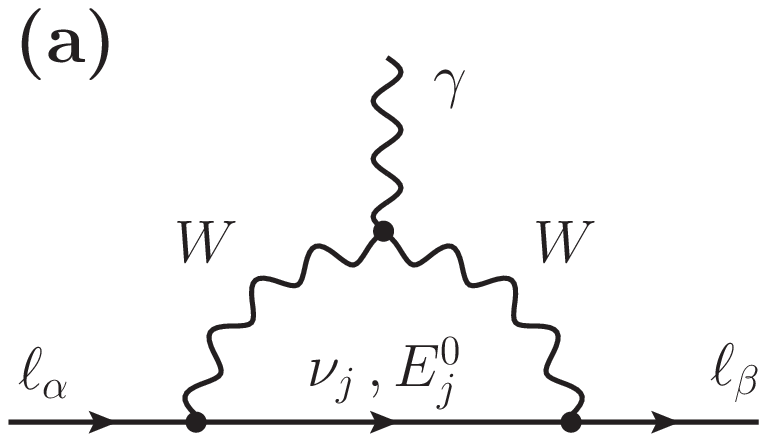}
\includegraphics[width=0.25\columnwidth]{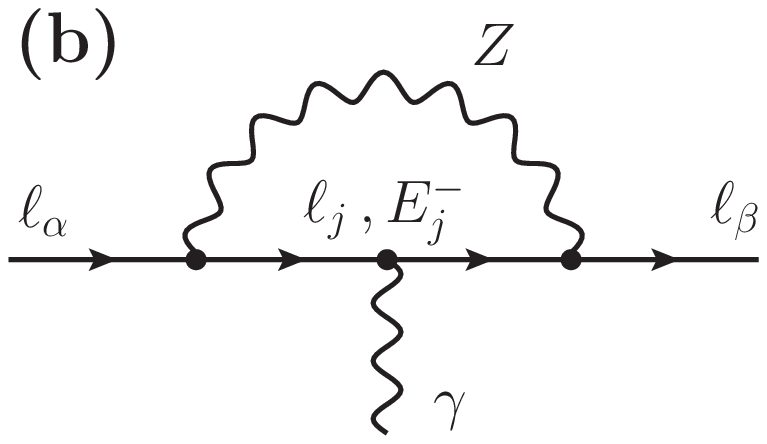}
\includegraphics[width=0.25\columnwidth]{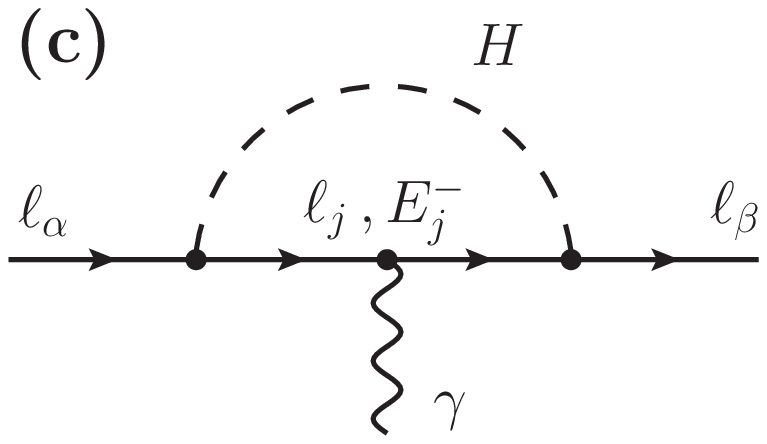}
\caption{Lowest-order diagrams that are relevant for the amplitude calculations of LFV decays ($\ell_\alpha \rightarrow \ell_\beta \gamma$) and anomalous magnetic moments of SM leptons (when $\alpha = \beta$) in the unitary gauge. Subscript $j$ denotes the flavor of the internal leptons, and is summed over in the computation. (a) The case mediated by $\nu_j$ corresponds to the usual diagram studied in standard electroweak theory, while the $E^0_j$ diagram comes from the new interactions; (b) \& (c) new contributing diagrams involving the $Z$ boson and physical Higgs, $H$.}
\label{fig:WZH_loop}
\end{center}
\end{figure}

Generically (because of gauge invariance), the transition amplitude for $\ell_\alpha \rightarrow \ell_\beta \gamma$ is given by the dimension-5 operator of the form
\begin{equation}\label{eqn:generic_EMDM_term}
 T(\ell_\alpha \rightarrow \ell_\beta\gamma) =
  \overline{u}_\beta \,
 (A+B\gamma_5)
  \, i \sigma_{\rho\nu}  q^\nu \varepsilon^\rho u_\alpha\;, \qquad \sigma_{\rho\nu} \equiv i \left[\gamma_\rho, \gamma_\nu\right]/2\;,
\end{equation}
where $A$ and $B$ correspond to the transition magnetic and electric dipole form factors\footnote{It is understood that $A$ and $B$ are dimensionful quantities when written in this form. Also, we have absorbed the extra factor of $i$ which is usually associated with the definition of the electric dipole moment into $B$.} respectively. In writing this down, we have used the on-shell condition, $q^2=0$, and $\varepsilon\cdot q =0$, where $q^\nu$ and $\varepsilon^\rho$ denote respectively, the photon 4-momentum and  polarization. 
In the SM with neutrino masses ($m_\nu > 0$), it is well-known that electroweak interactions involving the $W$ bosons in a loop (see Fig.~\ref{fig:WZH_loop}a) can give rise to a finite value for this amplitude although its size turns out to be vanishingly small because $M_W \gg m_\nu$ \cite{Cheng:1985bj, mu2eg_SM}. However, the situation may change drastically when there are new couplings to the SM leptons, such as those involving the exotic triplets studied here.

From Lagrangians (\ref{equ:L_CC}) to (\ref{equ:L_higgs}), we can identify all the new interactions and subsequently calculate the corresponding loop amplitudes from the definitions of
the modified coupling matrices given in (\ref{equ:g_LCC}) to (\ref{equ:g_H_both}). Working in the unitary gauge where only diagrams associated with the physical degrees of freedom are relevant, there are three types of one-loop graphs which may contribute to the LFV $\ell_\alpha \rightarrow \ell_\beta \gamma$ process (see Fig.~\ref{fig:WZH_loop}). As a result of the direct involvement of the triplet particles $E^0$ and $E^-$ in these diagrams, stringent constraints on $|\lambda_{\alpha\beta}|$ can be derived. These expressions are particularly useful when the expected improvement in experimental bounds are realized in the near future.

In calculating the amplitude for the lowest-order graphs in Fig.~\ref{fig:WZH_loop}, we note that any terms in (\ref{eqn:generic_EMDM_term}) that is proportional to $\overline{u}_\beta \gamma_\rho u_\alpha$ (or $\overline{u}_\beta \gamma_5\gamma_\rho u_\alpha$) will not contribute to the final answer as they cannot be transformed into the electromagnetic moment form \cite{Cheng:1985bj}. We can separate out this unwanted component from (\ref{eqn:generic_EMDM_term}) using Gordon identity, and get
\begin{equation}\label{eqn:generic_EMDM_gordon}
 T(\ell_\alpha \rightarrow \ell_\beta\gamma) =
  \overline{u}_\beta (A+B\gamma_5)
   (2p\cdot\varepsilon - m_\alpha \slashed{\varepsilon})u_\alpha
   - \overline{u}_\beta (A-B\gamma_5) m_\beta \slashed{\varepsilon}u_\alpha
   \;,
\end{equation}
where we have again used $\varepsilon\cdot q =0$ when simplifying the expression. In (\ref{eqn:generic_EMDM_gordon}), $p$ is the momentum of $\ell_\alpha$ while $m_{\alpha,\beta}$ denotes the $\ell_{\alpha,\beta}$ mass. Working in the limit where the final state lepton is assumed to be massless ($m_\beta\rightarrow 0$), one finds that amplitudes $A$ and $B$ become identical to leading order in $\lambda$, and thus in the explicit computation, we simply require to evaluate the coefficient of the $\overline{u}_\beta (1+\gamma_5)(2p\cdot \varepsilon)u_\alpha$ terms for all graphs.

Because we wish to work in the unitary gauge where there are less diagrams to consider,\footnote{One drawback is that some of the intermediate expressions/steps would be considerably more complicated than in other approaches.} our strategy is to perform the calculations in the notations of the generalized renormalizable ($R_\xi$) gauge \cite{Fujikawa:1972fe}, and at the end of the computation, we take the limit $\xi \rightarrow \infty$ to obtain the desired results.\footnote{We have adopted the definition of $\xi$ as used in modern textbooks \cite{Cheng:1985bj, Peskin:1995ev}, which is equivalent to the parameter $1/\xi$ as appeared in \cite{Fujikawa:1972fe}.} Moreover, we will work exclusively in the $m_{\ell_j} \ll M_{W,Z,H}$ and $m_\beta \ll 1$ limits (where $m_{\ell_j}$ represents the mass of the internal $j$-flavor SM lepton) and will only keep the leading order terms. 

After the dust has settled, we obtain the following expressions for the amplitudes of the one-loop contributions shown in Fig.~\ref{fig:WZH_loop} (superscripts and subscripts denote the type of internal leptons and bosons involved respectively):
\begin{align}
A_W^{\nu} &= \frac{-i G_F m_\alpha e}{8\pi^2\sqrt{2}} \left(
 -\frac{5}{3}\,\lambda_{\beta\alpha} -\sum_j \frac{m_{\nu_j}^2}{4M_W^2} \left(U_\nu\right)_{\beta j}
 \left(U_\nu^\dagger\right)_{j\alpha}
\right), \label{eqn:LFV_amp_W_nu}
\\
A_W^{E^0} &= \frac{-i G_F m_\alpha e}{8\pi^2\sqrt{2}}
 \sum_j \frac{v^2}{8} \left(Y_E M_E^{-1}\right)_{\beta j} \left(M_E^{-1} Y_E^\dagger \right)_{j\alpha}
  \left[f_1(w_j) + f_2(w_j)\right]\,, \qquad w_j \equiv M_{E_j}^2/M_W^2\;, 
\label{eqn:LFV_amp_W_E}
\\
A_Z^{\ell} &= \frac{-i G_F m_\alpha e}{8\pi^2\sqrt{2}} \,\frac{4(1+\sin^2\theta_w)}{3}\, \lambda_{\beta\alpha}\;,
\label{eqn:LFV_amp_Z_l}
\\
A_Z^{E^-} &= \frac{-i G_F m_\alpha e}{8\pi^2\sqrt{2}}
 \sum_j \frac{v^2}{8} \left(Y_E M_E^{-1}\right)_{\beta j} \left(M_E^{-1} Y_E^\dagger \right)_{j\alpha}
  \left[f_3(z_j) + f_4(z_j)+ f_5(z_j)\right]\,, \qquad z_j \equiv M_{E_j}^2/M_Z^2\;, 
\label{eqn:LFV_amp_Z_E}
\end{align}
\begin{align}
A_H^{\ell}&= \frac{-i G_F m_\alpha e}{8\pi^2\sqrt{2}}
 \sum_j \frac{3m_{\ell_j}^2}{2M_H^2} \ln \left(\frac{m_{\ell_j}^2}{M_H^2}\right) 
 \lambda_{\beta\alpha}\;,
\label{eqn:LFV_amp_H_l}
\\
A_H^{E^-} &= \frac{-i G_F m_\alpha e}{8\pi^2\sqrt{2}}
 \sum_j \frac{v^2}{8} \left(Y_E\right)_{\beta j} M_{E_j}^{-2} \left(Y_E^\dagger\right)_{j\alpha}
  \left[-2 f_5(h_j) +f_6(h_j) \right]\,, \qquad h_j \equiv M_{E_j}^2/M_H^2\;, 
\label{eqn:LFV_amp_H_E}
\end{align}
with
\begin{align}
 f_1(x) &= \frac{-10+43x -78x^2 +49x^3 -4x^4 -18x^3 \ln x}{12 (x-1)^4} \;,
 \label{eqn:f_1}
\\
 f_2(x) &= \frac{-4 +15x -12x^2 +x^3 +6x^2 \ln x}{2 (x-1)^3} \;,
 \label{eqn:f_2}
\\
 f_3(x) &= \frac{-4+9x -5x^3+ 6x(2x-1)\ln x}{3 (x-1)^4} \;,
 \label{eqn:f_3}
\\
 f_4(x) &= \frac{11x -18x^2 + 9x^3 -2x^4 +6x \ln x}{12 (x-1)^4} \;,
 \label{eqn:f_4}
\\
 f_5(x) &= \frac{-3x +4x^2 -x^3  -2x \ln x}{4 (x-1)^3} \;,
 \label{eqn:f_5}
\\
 f_6(x) &= \frac{2x +3x^2 -6x^3+ x^4 +6x^2 \ln x}{12 (x-1)^4} \;.
 \label{eqn:f_6}
\end{align}
In the above, $m_{\nu_j}$, $m_{\ell_j}$ and $M_{E_j}$ denote, respectively, the mass of the $j$-flavor neutrino, SM lepton and exotic $E$ particle. Note that the second term in (\ref{eqn:LFV_amp_W_nu}) is nothing but the usual contribution from neutrino mixing in standard electroweak theory \cite{Cheng:1985bj, Bilenky:1987ty}, where the $U_\nu$ matrix is the same as the one appeared in (\ref{eqn:N_defn}). 

So, we have for the total decay rate:
\begin{equation}
 \drate{\ell_\alpha \rightarrow \ell_\beta \gamma} = \frac{m_\alpha^3}{4\pi}
 \left|A_W^{\nu}+ A_W^{E^0} +A_Z^{\ell}+A_Z^{E^-}+A_H^{\ell} +A_H^{E^-}\right|^2 \;,
 \label{eqn:mu2g_rate}
\end{equation}
and subsequently, the branching ratio
\begin{align}
 \BR{\ell_\alpha \rightarrow \ell_\beta \gamma} &=  
  \frac{3\alpha_e}{2\pi} \left| \left[-\frac{1}{3} +\frac{4}{3} \sin^2\theta_w +\sum_j \frac{3m_{\ell_j}^2}{2M_H^2} \ln \left(\frac{m_{\ell_j}^2}{M_H^2}\right) \right] \lambda_{\beta\alpha} 
    -\sum_j \frac{m_{\nu_j}^2}{4M_W^2} \left(U_\nu\right)_{\beta j}
 \left(U_\nu^\dagger\right)_{j\alpha} 
 \right. \nonumber\\
 &\qquad\qquad
 + \sum_j \frac{v^2}{8} \left(Y_E M_E^{-1}\right)_{\beta j} \left(M_E^{-1} Y_E^\dagger \right)_{j\alpha}
  \left[f_1(w_j) + f_2(w_j) + f_3(z_j) + f_4(z_j)+ f_5(z_j)\right]
     \nonumber\\
 &\qquad\qquad
 \left.+ \sum_j \frac{v^2}{8} \left(Y_E\right)_{\beta j} M_{E_j}^{-2} \left(Y_E^\dagger\right)_{j\alpha}
  \left[-2 f_5(h_j) +f_6(h_j) \right]
 \right|^2
 \BR{\ell_\alpha \rightarrow \ell_\beta \nu_\alpha \overline{\nu}_\beta}\;,
\label{eqn:mu2g_BR} 
\end{align}
where $\alpha_e$ is the fine-structure constant.
Taking $M_{E_j} \simeq 100$~GeV (the lower bound for heavy charged leptons \cite{Nakamura:2010zzi}) for all $j$, and assuming the Higgs mass, $M_H$, is about 114~GeV \cite{Barate:2003sz}, the experimental limits \cite{Nakamura:2010zzi} on \BR{\mu\rightarrow e\gamma}, \BR{\tau\rightarrow e\gamma} and \BR{\tau\rightarrow \mu\gamma} then lead to\footnote{Note that if we have chosen a larger value for the mass of $E_j$, it will only make these bounds less stringent.}
\begin{align}
 &|\lambda_{e\mu}|< 2.2 \times 10^{-4}\;, \label{eqn:Lee_bound_loop}\\
 &|\lambda_{e\tau}|< 2.7 \times 10^{-2}\;, \label{eqn:Lee_bound_loop}\\
&|\lambda_{\mu\tau}|< 3.1 \times 10^{-2}\; \label{eqn:Lee_bound_loop}.
\end{align}
These bounds are not as strong as those displayed in (\ref{eqn:Lee_bound_tree}) to (\ref{eqn:Lmt_bound_tree}) which come from tree-level interactions. However, improvement is expected when the new and ongoing experiments mentioned have reached their projected sensitivities.

\subsection{$\mu$-$e$ conversion in atomic nuclei}\label{subsec:m_to_e}

Owing to the fact that coherent contribution of all nucleons in the nucleus can enhance the experimental signals, muon-to-electron conversion in muonic atoms provides another excellent platform for studying tree-level FCNC. Not only does it place a constraint on the same $\mu$-$e$-$Z$ vertex as appeared in $Z\rightarrow e^\pm\mu^\mp$, $\mu\rightarrow 3e$ and the loop graphs in Fig~\ref{fig:WZH_loop}b, as we shall show below, its bound on $|\lambda_{e\mu}|$ is the most stringent amongst all applicable LFV interactions considered. The test for $\mu$-$e$ conversion, therefore, plays a complementary role to the investigation of $\mu\rightarrow e\gamma$ in the probe for physics beyond the SM as they are induced differently. 

In what follows, we shall assume that the only contribution to the $\mu$-$e$ conversion rate in our setup comes from exchanges with the $Z$ bosons. This approximation is sensible because the cases mediated by the photon and the Higgs are suppressed by loop effects and $M_H^{-1}$ respectively. So, at the quark level and after integrating out $M_Z$, the effective interaction Lagrangian which can induce the $\mu$-$e$ transition can be written as\footnote{We have assumed that standard electroweak interaction operates in the quark sector.}
\begin{align}
 \mathcal{L}^\text{eff}_{\mu\rightarrow e}
  &= \sqrt{2} \, G_F \overline{\ell}_e \gamma^\nu (k_V - k_A \gamma_5) \ell_\mu 
  \left[
  \overline{q}_u \gamma_\nu (v_u + a_u \gamma_5) q_u
  +
  \overline{q}_d \gamma_\nu (v_d + a_d \gamma_5) q_d
  \right] \;, \label{eqn:m2e_L_eff}
\end{align}
where $q_{u,d}$ denotes the $u, d$-quark field while
\begin{eqnarray}
 k_V = k_A = -\lambda_{e\mu}\;, \quad
 a_u = -a_d = -\frac{1}{2}\;,\quad
 v_u = \frac{1}{2}-\frac{4}{3} \sin^2 \theta_w \;, \quad
 v_d = -\frac{1}{2}+\frac{2}{3} \sin^2 \theta_w \;.
\end{eqnarray}

Appealing to the general result obtained from FCNC analysis with massive gauge bosons in \cite{Bernabeu:1993ta}, we then arrive at the following expression for the branching ratio of $\mu$-$e$ conversion in nuclei:\footnote{This result is a good approximation only for nuclei with less than about 100 nucleons.}
\begin{equation}\label{eqn:m2e_BR_formula}
  B_{\mu\rightarrow e} \simeq 
  \frac{G_F^2 \,\alpha_e^3 \,m_\mu^3\, p_e'\, E_e'}{\pi^2\,\Gamma_\text{cap}^\mathcal{A}} 
  \left|F(q'^2)\right|^2
  \left(k_V^2+k_A^2\right) \frac{\mathcal{Z}_\text{eff}^4\, \hat{Q}^2}{\mathcal{Z}} \;,
\end{equation}
where $p_e'$ ($E_e'$) is the momentum (energy) of the electron, $\Gamma_\text{cap}^\mathcal{A}$ represents the total nuclear muon capture rate for element $\mathcal{A}$, and 
$\mathcal{Z}$ ($\mathcal{Z}_\text{eff}$) is the (effective) atomic number of the element under investigation. In (\ref{eqn:m2e_BR_formula}),
$F(q'^2)$ is the nuclear form factor which may be measured from electron scattering experiments \cite{11of_ref22} while %
\begin{equation}
 \hat{Q} = (2\mathcal{Z} + \mathcal{N}) \, v_u + (\mathcal{Z} + 2\mathcal{N})\, v_d \;,
\end{equation}
with $\mathcal{N}$ denoting the number of neutrons in the nuclei.

Given that one of the best upper limit on the $\mu$-$e$ conversion branching ratio is obtained from measurements with titanium-48 ($^{48}_{22}\text{Ti}$) in the SINDRUM II experiments \cite{Dohmen:1993mp}: 
\begin{equation}\label{eqn:m2e_BR_Ti}
 B_{\mu\rightarrow e}^\text{exp} \equiv 
 \frac{\drate{\mu^-\,\text{Ti} \rightarrow e^-\,\text{Ti}}}{\Gamma_\text{cap}^\text{Ti}} 
 < 4.3 \times 10^{-12}\;,
\end{equation}
we shall use the parameters for element $^{48}_{22}\text{Ti}$ in (\ref{eqn:m2e_BR_formula}) to deduce our bound.\footnote{Although the value quoted in the experiments with gold (Au): 
$\drate{\mu^-\text{Au} \rightarrow e^-\text{Au}} / \Gamma_\text{cap}^\text{Au} <
7\times 10^{-13}$ \cite{Bertl:2006up} is smaller than the one in (\ref{eqn:m2e_BR_Ti}), theoretical calculations \cite{Kitano:2002mt} have shown that for very heavy elements (atomic number $\mathcal{Z} \gtrsim 60$) like Au, the $\mu$-$e$ conversion rate is actually suppressed. Therefore, this does not necessarily indicate a better bound on the rate, especially when the estimation of the nuclear matrix element for such heavy nuclei can carry large uncertainties.}
Following the approximation as applied in \cite{Bernabeu:1993ta}, we take $p_e'\simeq E_e' \simeq m_\mu$, and $F(q'^2 \simeq -m_\mu^2) \simeq 0.54$. In addition, we have $\mathcal{Z}_\text{eff} \simeq 17.6$ for $^{48}_{22}\text{Ti}$ \cite{Zeff} and $\Gamma_\text{cap}^\mathcal{A} \equiv \Gamma_\text{cap}^\text{Ti} \simeq 2.59\times 10^6$~s$^{-1}$ \cite{Suzuki:1987jf}. Hence, (\ref{eqn:m2e_BR_formula}) and (\ref{eqn:m2e_BR_Ti}) combine to give
\begin{equation}\label{eqn:m2e_Lem_bound}
 |\lambda_{e\mu}| < 5.3 \times 10^{-7}\;,
\end{equation}
where in the derivation, we have taken into account the modification of $G_F$ as discussed in (\ref{eqn:modified_GF}) and subsequently, substituted in the values from (\ref{eqn:Laa_values}). 

As foreshadowed, the bound displayed in (\ref{eqn:m2e_Lem_bound}) is indeed the most stringent one on $|\lambda_{e\mu}|$, and with the new experiments being planned, respectively, at J-PARC and Fermilab by the COMET (and PRISM/PRIME) \cite{exp_COMET} and Mu2e \cite{exp_mu2e} collaborations, it is expected to become even stronger in the near future.

\section{Global fit on the elements of $\lambda$ and some consequences}\label{sec:global_fit}

In this section, we bring together all the results obtained thus far and perform a global analysis on the
elements of the $\lambda$ matrix, which are key to determining the new physics effects from the triplet leptons. For convenience, a summary of all constraints derived in Sec.~\ref{sec:W_Z_decays} and \ref{sec:LFV_all} are collected in Table~\ref{table:all_results}.

\begin{table}[ht]
\begin{center}
\begin{tabular}{|c|c|c|c|}
\hline
parameter(s) & process & limit on BR & constraint on $\lambda$'s\\
\hline
$\lambda_{ee}$ & $Z\rightarrow e^- e^+$ &$3.363\pm 0.004 \times 10^{-2}$&  $\; -2.7 \mp 0.4 \times 10^{-3}\; $\\
$\lambda_{\mu\mu}$ & $Z\rightarrow \mu^- \mu^+$ &$3.366\pm 0.007 \times 10^{-2}$&  $\; -2.9 \mp 0.4 \times 10^{-3}\; $\\
$\lambda_{\tau\tau}$ & $Z\rightarrow \tau^- \tau^+$ &$3.369\pm 0.008 \times 10^{-2}$&  $\; -3.1 \mp 0.4 \times 10^{-3}\; $\\
\hline
$|\lambda_{e\mu}|$ & 
\begin{tabular}{c}
 $Z\rightarrow e^{\pm}\mu^{\mp}$\\ 
 $\mu^- \rightarrow e^- e^- e^+$\\
 $\mu\rightarrow e \gamma$\\
 $\mu$-$e$ conversion\\
\end{tabular}   
&
\begin{tabular}{c}
 $<1.7 \times 10^{-6}$\\ 
 $<1.0 \times 10^{-12}$\\
 $<1.2 \times 10^{-11}$\\
 $<4.3 \times 10^{-12}$ (Ti)\\
\end{tabular}  
&
\begin{tabular}{c}
 $< 1.8 \times 10^{-3}$  \\
 $< 1.1 \times 10^{-6}$  \\
 $< 2.2 \times 10^{-4}$  \\ 
 $< 5.3 \times 10^{-7}$  \\
\end{tabular} \\
\hline
$|\lambda_{e\tau}|$ & 
\begin{tabular}{c}
 $Z\rightarrow e^{\pm}\tau^{\mp}$\\ 
 $\tau^- \rightarrow e^- e^- e^+$\\
 $\tau^- \rightarrow e^- \mu^- \mu^+$\\
 $\tau\rightarrow e \gamma$\\
\end{tabular}   
& 
\begin{tabular}{c}
 $<9.8 \times 10^{-6}$\\ 
 $<3.6 \times 10^{-8}$\\
 $<3.7 \times 10^{-8}$\\
 $<3.3 \times 10^{-8}$\\
\end{tabular}  
&
\begin{tabular}{c}
 $< 4.3 \times 10^{-3}$  \\
 $< 5.0 \times 10^{-4}$  \\
 $< 6.5 \times 10^{-4}$  \\ 
 $< 2.7 \times 10^{-2}$  \\
\end{tabular}   
\\ 
\hline
$|\lambda_{\mu\tau}|$ & 
\begin{tabular}{c}
 $Z\rightarrow \mu^{\pm}\tau^{\mp}$\\ 
 $\tau^- \rightarrow \mu^- \mu^- \mu^+$\\
 $\tau^- \rightarrow \mu^- e^- e^+$\\
 $\tau\rightarrow \mu \gamma$\\
\end{tabular}   
&
\begin{tabular}{c}
 $<1.2 \times 10^{-5}$\\ 
 $<3.2 \times 10^{-8}$\\
 $<2.7 \times 10^{-8}$\\
 $<4.4 \times 10^{-8}$\\
\end{tabular}  
& 
\begin{tabular}{c}
 $< 4.7 \times 10^{-3}$  \\
 $< 4.8 \times 10^{-4}$  \\
 $< 5.5 \times 10^{-4}$  \\ 
 $< 3.1 \times 10^{-2}$  \\
\end{tabular}   
\\ 
\hline
$|\lambda_{\mu e}||\lambda_{\mu\tau}|$ & $\tau\rightarrow e^+ \mu^-\mu^-$ &$<2.3 \times 10^{-8}$&  $< 2.6 \times 10^{-4} $\\
$|\lambda_{e\mu}||\lambda_{e\tau}|$ & $\tau\rightarrow \mu^+ e^- e^-$ &$<2.0 \times 10^{-8}$&  $< 2.4 \times 10^{-4}$\\
\hline
\end{tabular}
\caption{A collection of all constraints on the elements of $\lambda \equiv v^2 Y_E M_E^{-2} Y_E^\dagger/8$ from processes studied in Sec.~\ref{sec:W_Z_decays} and \ref{sec:LFV_all}.}\label{table:all_results}
\end{center}
\end{table}

Studying the results listed in Table~\ref{table:all_results} and recalling that $|\lambda_{\alpha\beta}| =|\lambda_{\beta\alpha}|$, it is not difficult to obtain the following overall fit for the elements of $\lambda$: 
\begin{equation}\label{eqn:global_constraint}
 \ththMat{|\lambda_{ee}|}{|\lambda_{e\mu}|}{|\lambda_{e\tau}|}
 {|\lambda_{\mu e}|}{|\lambda_{\mu \mu}|}{|\lambda_{\mu \tau}|}
 {|\lambda_{\tau e}|}{|\lambda_{\tau \mu}|}{|\lambda_{\tau \tau}|}
 \lesssim
 \ththMat{2.7 \times 10^{-3}}{< 5.3 \times 10^{-7}}{< 5.0 \times 10^{-4}}
 {< 5.3 \times 10^{-7}}{2.9 \times 10^{-3}}{< 4.8 \times 10^{-4}}
 {< 5.0 \times 10^{-4}}{< 4.8 \times 10^{-4}}{3.1 \times 10^{-3}}\;.
\end{equation}
Relation (\ref{eqn:global_constraint}) is one of our major results in this work. Note that the off-diagonal elements were derived assuming $M_E \simeq 100$~GeV (for all flavors). So, this also provides a rough estimate of the size for the exotic coupling $Y_E$ (in this interesting mass range for $M_E$):
\begin{equation}
 \left|Y_E\right|_{\alpha\beta} \lesssim \order{10^{-2}}\text{ to } \order{10^{-3}}\;, \qquad
 \text{for all } \alpha, \beta\;.
\end{equation}
Should $M_E$ be any heavier the bounds in (\ref{eqn:global_constraint}) are expected to loosen and the corresponding range allowed for $|Y_E|_{\alpha\beta}$ would be increased.

Moreover, if one assumes that the nonzero elements of $\lambda$ induced by mixing with the exotic triplets are the \emph{only} source of LFV in the theory, one may relate the various branching ratios discussed above as follows:
\begin{align}
 \BR{Z\rightarrow e^{\pm}\mu^{\mp}} &\simeq 3.6\times 10^{-2} \;\BR{\mu\text{-}e \,\text{conversion in Ti}}\;, \nonumber\\
 \BR{\mu^- \rightarrow e^- e^- e^+} &\simeq 5.3\times 10^{-2} \;\BR{\mu\text{-}e \,\text{conversion in Ti}}\;, \nonumber\\
 \BR{\mu \rightarrow e \gamma} &\simeq 1.7\times 10^{-5} \;\BR{\mu\text{-}e \,\text{conversion in Ti}}\;,
\label{eqn:BR_relate_em}
\end{align}
for the processes involving $|\lambda_{e\mu}|$. Whereas for $|\lambda_{e\tau}|$ and  $|\lambda_{\mu\tau}|$, one gets 
\begin{align}
 \BR{Z\rightarrow e^{\pm}\tau^{\mp}} &\simeq 3.8\times 10^{0} \; \BR{\tau^- \rightarrow e^- e^- e^+}\;, \nonumber\\
  \BR{\tau^- \rightarrow e^- \mu^- \mu^+} &\simeq 6.2\times 10^{-1} \; \BR{\tau^- \rightarrow e^- e^- e^+}\;, \nonumber\\
 \BR{\tau \rightarrow e \gamma} &\simeq 3.3\times 10^{-4} \; \BR{\tau^- \rightarrow e^- e^- e^+}\;,
\label{eqn:BR_relate_et}
\intertext{and}
 \BR{Z\rightarrow \mu^{\pm}\tau^{\mp}} &\simeq 3.9\times 10^{0} \; \BR{\tau^- \rightarrow \mu^- \mu^- \mu^+}\;, \nonumber\\
  \BR{\tau^- \rightarrow \mu^- e^- e^+} &\simeq 6.5\times 10^{-1} \; \BR{\tau^- \rightarrow \mu^- \mu^- \mu^+}\;, \nonumber\\
 \BR{\tau \rightarrow \mu \gamma} &\simeq 3.3\times 10^{-4} \; \BR{\tau^- \rightarrow \mu^- \mu^- \mu^+}\;,
\label{eqn:BR_relate_mt}
\end{align}
respectively. In the above, we have again taken $M_E \simeq 100$~GeV.\footnote{We have checked that taking $M_E \rightarrow \infty$ would only change the numerical factor by an insignificant amount of at most 30\%.} 

Applying the experimental limits on the right-hand side of relations (\ref{eqn:BR_relate_em}) to (\ref{eqn:BR_relate_mt}), we can obtain the model specific bounds on the branching ratio of many key LFV processes.
For instance, (\ref{eqn:BR_relate_em}) implies $\BR{\mu \rightarrow e \gamma} \lesssim 7.3 \times 10^{-17}$ in this theory with exotic triplets. Notice that this is significantly stronger than the current limit set by experiments. As a result, a future detection of this LFV process above this rate will invalidate the predictions of this minimal extension to the SM, and point to the existence of other new physics in the lepton sector. 
Similar conclusions may also be drawn from other processes displayed above.

\section{Contribution to lepton anomalous magnetic moment}\label{sec:AMM}

While in Dirac theory the gyromagnetic ratio of a spin-1/2 particle is predicted to have a value of $\tilde{g}^\text{dirac} = 2$, it is well-known that quantum field theory gives a correction to this number via loop effects. 
The deviation from the Dirac result of 2  is usually parameterized by the dimensionless quantity ($\alpha$ denotes the flavor)
\begin{equation} \label{eqn:g-2_defn}
 a_\alpha \equiv \frac{\tilde{g}_\alpha - 2}{2}\;,\quad \text{where $\tilde{g}_\alpha$ is the actual value of the gyromagnetic ratio,}
\end{equation}
known as the \emph{anomalous magnetic moment}. It is related to the lepton magnetic dipole moment $\vec{\mu}_\alpha = -e(1+a_\alpha)/2m_\alpha \,\vec s$, where $\vec s$ is the unit spin vector. In terms of the parameters from quantum field theory, $a_\alpha \equiv F_2(q^2=0)$, when the form factor expansion for a general lepton-photon amplitude is written as
\begin{equation}\label{eqn:form_factors_expand}
 T(\ell_\alpha \rightarrow \ell_\alpha'\gamma) =
 -i e\, \overline{u}_\alpha' 
 \left[ F_1(q^2) \gamma_\rho
  +\frac{F_2(q^2)}{2m_\alpha}\,i\sigma_{\rho\nu} \,q^\nu
  +\frac{F_3(q^2)}{2m_\alpha}\, \sigma_{\rho\nu} \gamma_5 \,q^\nu +\cdots
 \right]
  \varepsilon^\rho u_\alpha
   \;, \quad e>0\;,
\end{equation}
where $q^\nu$ is again the photon momentum (see (\ref{eqn:generic_EMDM_term}) for notations).\footnote{Note that the lepton electric dipole moment is proportional to $F_3(q^2=0)$.} Therefore, the precise contribution to $a_\alpha$ from the SM (and indeed any other theories) can be calculated by considering all the relevant loop diagrams for the $F_2(0)$-term. 

While the anomalous magnetic moment for the electron, muon and tauon can all be very important in their own rights, given the present experimental and theoretical development, $a_\mu$ is the most interesting observable to examine. This is because when combining the fact that significant contributions to the overall predicted $a_\mu$ value come from every major sectors (QED, electroweak, hadronic) of the SM \cite{g-2_review1, mg-2_SM} with the ability to experimentally measure $a_\mu$ to extremely high accuracy \cite{mg-2_exp1, mg-2_exp2}, the SM as a whole can be scrutinized, and any discrepancies between theory and experiment would be a strong indication of new physics. On the other hand, although $a_e$ have been measured to extraordinary precision (hence providing a very stringent test on QED and the value of the fine-structure constant~$\alpha_e$ \cite{eg-2_exp, eg-2_fineS}), its low sensitivity to the contributions from strong and electroweak processes means that any hypothetical modifications to these sectors (due to new physics) would not be easily detectable. As far as $a_\tau$ is concerned, even though its much heavier mass would in theory imply better sensitivity to any new physics than $a_\mu$, its usefulness has been limited by the relatively poor experimental bounds. In fact, the best current limits set by the DELPHI experiments \cite{DELPHI} are still too coarse to even check the first significant figure of $a_\tau$ from theoretical calculations.

Currently, the experimental values for $a_e$~\cite{eg-2_exp}, $a_\mu$~\cite{mg-2_exp2} and $a_\tau$~\cite{DELPHI} are given by 
\begin{align}
 a_e^\text{Exp} &= 115~965~218~073(28)\times 10^{-14}\;, \label{eqn:a_e_exp}
 \\
 a_\mu^\text{Exp} &= 116~592~089(63)\times 10^{-11}\;, \label{eqn:a_mu_exp} 
 \\
 a_\tau^\text{Exp} &= 
 \begin{cases}
 \; < 1.3 \times 10^{-2} \;,\\ 
 \; > -5.2 \times 10^{-2} \;.
 \end{cases}
 \label{eqn:a_tau_exp} 
\end{align}
Focusing on the muon case, one finds that the discrepancy between experiment and the SM estimate is about 4.0$\sigma$ \cite{mg-2_SM}:
\begin{equation} \label{eqn:mg-2_diff}
 \Delta a_\mu = a_\mu^\text{Exp} - a_\mu^\text{SM} = 316(79)\times 10^{-11}\;.
\end{equation}
If this difference is real (rather than caused by incorrect leading-order hadronic approximation\footnote{Although this possibility is not completely ruled out, shifting the hadronic cross-section to bridge this gap will naturally increase the tension with the lower bound on the Higgs mass, both from LEP \cite{Barate:2003sz}
and the SM vacuum stability requirement \cite{mg-2_SM}.}), then there must be some new physics at play. In the following, we investigate whether the effects induced by the exotic triplets can have an influential role on this front.

The procedure for calculating the anomalous magnetic moment due to the modified electroweak couplings of (\ref{equ:g_LCC}) to (\ref{equ:g_H_both}) is in fact analogous to the computation for LFV $\ell_\alpha\rightarrow\ell_\beta\gamma$ done in Sec.~\ref{subsec:l_lg}. Working in the unitary gauge again, the one-loop diagrams one needs to consider are the three main types depicted in Fig.~\ref{fig:WZH_loop}, but with the condition $\alpha=\beta$ imposed. As a result, the approximation of massless final state lepton cannot be used any more (i.e. $m_\beta \not\rightarrow 0$). In addition, since we are only interested in the magnetic moment, the part associated with $\gamma_5$ in the general amplitude
\begin{equation}
 T(\ell_\alpha \rightarrow \ell_\alpha'\gamma) =
  \overline{u}_\alpha' \,
  (C+D\gamma_5)
  \,i \sigma_{\rho\nu}  q^\nu \varepsilon^\rho u_\alpha\;, \label{eqn:gen_mag_moment}
\end{equation}
will be disregarded. Hence, in the computation, the terms to concentrate on are those proportional to $\overline{u}_\alpha' (2p\cdot \varepsilon)u_\alpha$, where $p$ is the momentum of the incoming $\ell_\alpha$.
Apart from these changes, the general strategy is identical to Sec.~\ref{subsec:l_lg}.

Employing a similar notation system as before, the amplitudes of the one-loop diagrams from Fig.~\ref{fig:WZH_loop} for the case $\alpha = \beta$ are given by (to leading order)
\begin{align}
 C_W^{\nu} &= \frac{-i G_F m_\alpha e}{8\pi^2\sqrt{2}}
  \left(\frac{5}{3} -\frac{10}{3}\,\lambda_{\alpha\alpha} \right), \label{eqn:AMM_W_nu}
 \\
 C_W^{E^0} &= \frac{-i G_F m_\alpha e}{8\pi^2\sqrt{2}}
  \sum_j \frac{v^2}{8} \left(Y_E M_E^{-1}\right)_{\alpha j} \left(M_E^{-1} Y_E^\dagger \right)_{j\alpha}
  \left[f_7(w_j) + 3 f_8(w_j) + f_9(w_j) +1 \right], \quad w_j \equiv M_{E_j}^2/M_W^2\;, 
 \label{eqn:AMM_W_E}
 \\
%
  C_Z^{\ell} &= \frac{-i G_F m_\alpha e}{8\pi^2\sqrt{2}}
  \left(
   \frac{8}{3} \sin^4\theta_w
  -\frac{4}{3} \sin^2\theta_w
  -\frac{2}{3}
  +\frac{8}{3} (1+ \sin^2\theta_w)\,\lambda_{\alpha\alpha}
  \right),
 \label{eqn:AMM_Z_l}
 \\
   C_Z^{E^-} &= \frac{-i G_F m_\alpha e}{8\pi^2\sqrt{2}}
  \sum_j \frac{v^2}{8} \left(Y_E M_E^{-1}\right)_{\alpha j} \left(M_E^{-1} Y_E^\dagger \right)_{j\alpha}
  \left[f_3(z_j) + 2 f_4(z_j) + 2f_5(z_j) \right], \quad z_j \equiv M_{E_j}^2/M_Z^2\;,
 \label{eqn:AMM_Z_E}
 \\
   C_H^{\ell} &= \frac{-i G_F m_\alpha e}{8\pi^2\sqrt{2}}\, (6\lambda_{\alpha\alpha} -1)\;
   \order{m_{\ell_j}^2/M_H^2} \simeq 0\;,
   \label{eqn:AMM_H_l}
   \\
 C_H^{E^-} &= \frac{-i G_F m_\alpha e}{8\pi^2\sqrt{2}}\,\frac{v^2}{8}
 \sum_j  \left(Y_E\right)_{\alpha j} M_{E_j}^{-2} \left(Y_E^\dagger\right)_{j\alpha}
  2 f_6(h_j) 
  \nonumber\\
  &\qquad\quad
  -\left[\left(Y_E M_E^{-1}\right)_{\alpha j} M_{E_j}^{-1} \left(Y_E^\dagger\right)_{j\alpha}
  +\left(Y_E\right)_{\alpha j} M_{E_j}^{-1} \left(M_{E}^{-1}Y_E^\dagger\right)_{j\alpha}
  \right] 2 f_5(h_j)   
  \,, \quad h_j \equiv M_{E_j}^2/M_H^2\;, 
\label{eqn:AMM_H_E}  
\end{align}
where $f_{3}(x)$ to $f_{6}(x)$ are given in (\ref{eqn:f_3}) to (\ref{eqn:f_6}) and
\begin{align}
 f_7(x) &= \frac{7-33x +57x^2 -31x^3 +6x^2(3x-1) \ln x}{6 (x-1)^4} \;,
 \label{eqn:f_7}
\\
 f_8(x) &= \frac{-1 +4x -3x^2 + 2 x^2 \ln x}{(x-1)^3} \;,
 \label{eqn:f_8}
\\
 f_9(x) &= \frac{3 -10x +21x^2 -18x^3+ 4x^4 +6x^2\ln x}{6 (x-1)^4} \;.
 \label{eqn:f_9}
\end{align}
Comparing (\ref{eqn:gen_mag_moment}) with the form factor expansion of (\ref{eqn:form_factors_expand}), the anomalous magnetic moment can be written in terms of the amplitudes computed above:
\begin{equation}\label{eqn:amp_FF_rel}
 a_\alpha \equiv F_2(0) = -\frac{2m_\alpha}{i e} \left(
 C_{W}^{\nu}+C_{W}^{E^0}+C_{Z}^{\ell}+C_{Z}^{E^-}+C_{H}^{\ell}+C_{H}^{E^-}
 \right)\;.
\end{equation}
Note that the result given in (\ref{eqn:amp_FF_rel}) contains the usual SM electroweak component of $a_\alpha$, as well as 
the contribution induced by the new physics. Examining our results, we see that the terms which are \emph{not} proportional to $\lambda_{\alpha\alpha}$ in (\ref{eqn:AMM_W_nu}) and (\ref{eqn:AMM_Z_l}) sum up to give the usual prediction of the anomalous magnetic moment from the SM \cite{Fujikawa:1972fe}. 
Removing this component from (\ref{eqn:amp_FF_rel}) and using the values for $\lambda_{\alpha\alpha}$ given in Table~\ref{table:all_results}, we obtain the following estimate for the anomalous magnetic moments coming from the new physics associated with the exotic lepton triplets:\footnote{These values are obtained assuming the contribution from (\ref{eqn:AMM_H_l}) is exactly zero. If one uses the full expression for (\ref{eqn:AMM_H_l}), the results would only shift by a numerical factor of \order{1}.}
\begin{align}
 \Delta a_e^E    &\simeq -5.4\times 10^{-16}\;, \label{eqn:a_e_result}\\
 \Delta a_\mu^E &\simeq -2.5\times 10^{-11}\;, \label{eqn:a_m_result}\\
 \Delta a_\tau^E &\simeq -7.6\times 10^{-9}\;, \label{eqn:a_t_result}
\end{align}
where we have again assumed $M_{E_j}\simeq 100$~GeV and $M_H \simeq 114$~GeV. Looking at (\ref{eqn:a_e_result}) to (\ref{eqn:a_t_result}), we note that these values are all \emph{negative}, meaning that they would not have been helpful in explaining the discrepancy between SM prediction and experiments even if they are of the correct magnitude.\footnote{Interestingly, a similar conclusion has been reached in type-III seesaw models \cite{Biggio:2008in}.} As it turns out, these contributions are at least one order of magnitude less than the experimental errors given for the quantities listed in (\ref{eqn:a_e_exp}) to (\ref{eqn:a_tau_exp}). Therefore, we do not expect the new physics effects from the exotic triplets to be distinguishable from the SM components in these experiments.

\section{Conclusion}

Given that the phenomenology of nonzero neutrino masses motivates an extension to the SM, it is natural to ask what might be the simplest ways that new physics can couple to known particles. Working exclusively in the lepton sector and demanding renormalizability and SM gauge invariance as the basic requirements, we concentrated on the minimal couplings where there is only one exotic particle appearing per term. Whilst we have found that the phenomenologies of the majority of exotic particles allowed by this framework have already been closely studied due to other motivations such as neutrino mass generation and grand unification, some interesting possibilities remain unexplored. Thus, in this work we have focused our attention on one of those, namely the exotic lepton triplets that carries nonzero hypercharge and can Yukawa couple to ordinary LH leptons. 

Using a formalism that is similar to that used in the analyses of seesaw models, we have identified and defined the key parameter (denoted $\lambda$ throughout) that encapsulates the new physics effects caused by the introduction of these exotic triplets. In particular, we note that the off-diagonal entries of this $\lambda$ matrix are the new sources for FCNC phenomenologies. By invoking the limits from low-energy experiments, constraints can then be placed on this dimensionless parameter that controls the coupling strength with the exotic leptons. Such investigation is particularly worthwhile given that these minimally coupled triplets may give rise to definite collider signatures at the LHC \cite{DelNobile:2009st}.

In this paper, we have studied the implications from leptonic $W$ and $Z$ decays and found that the diagonal elements, $|\lambda_{\alpha\alpha}|$, must be of \order{10^{-3}} to agree with the current limits, while bounds for the off-diagonal entries were obtained by investigating their effects on LFV processes such as $\ell\rightarrow 3\ell$, $\ell\rightarrow \ell'\gamma$ and $\mu$-$e$ conversion in titanium nuclei. With the exception of $|\lambda_{e\mu}|$ which has its strongest constraint coming from the $\mu$-$e$ conversion experiments, other off-diagonal values receive their most stringent bounds from LFV $\ell\rightarrow 3\ell$ decays. Some of these limits are expected to improve significantly when the next generation of experiments have reached their proposed sensitivity.

Furthermore, the contribution to the lepton anomalous magnetic moment from the new physics was investigated. Through explicit computation of the relevant lowest-order loop graphs, it has been concluded that any potential contributions on this front are far too small to be detected in experiments at the present time. As a result, introducing this type of exotic triplets of leptons into the SM cannot help to explain the muon $g-2$ anomaly.

\section*{Acknowledgements}

SSCL would like to thank A.~G.~Akeroyd for discussion. This work is supported in part by the NSC (grant numbers: NSC-97-2112-M-033-002-MY3 and NSC-99-2811-M-033-013) and in part by the NCTS of Taiwan.







\begin{thebibliography}{99}

\bibitem{neutrinos_exp}
  B.~T.~Cleveland {\it et al.},
  Nucl.\ Phys.\ Proc.\ Suppl.\  {\bf 38}, 47 (1995);
  Y.~Fukuda {\it et al.}  [Super-Kamiokande Collaboration],
  Phys.\ Rev.\ Lett.\  {\bf 82}, 2430 (1999)
  [arXiv:hep-ex/9812011];
   Y.~Fukuda {\it et al.}  [Super-Kamiokande Collaboration],
  Phys.\ Rev.\ Lett.\  {\bf 81}, 1562 (1998)
  [arXiv:hep-ex/9807003];
  K.~Lande {\it et al.},
  Nucl.\ Phys.\ Proc.\ Suppl.\  {\bf 77}, 13 (1999);
  D.~N.~Abdurashitov {\it et al.}  [SAGE Collaboration],
  Nucl.\ Phys.\ Proc.\ Suppl.\  {\bf 77}, 20 (1999);
  T.~A.~Kirsten  [GALLEX and GNO Collaborations],
  Nucl.\ Phys.\ Proc.\ Suppl.\  {\bf 77}, 26 (1999);
  Y.~Fukuda {\it et al.}  [Super-Kamiokande Collaboration],
  Phys.\ Lett.\ B {\bf 436}, 33 (1998)
  [arXiv:hep-ex/9805006];
  C.~Athanassopoulos {\it et al.}  [LSND Collaboration],
  Phys.\ Rev.\ C {\bf 54}, 2685 (1996)
  [arXiv:nucl-ex/9605001];
  C.~Athanassopoulos {\it et al.}  [LSND Collaboration],
  Phys.\ Rev.\ C {\bf 58}, 2489 (1998)
  [arXiv:nucl-ex/9706006];
  Q.~R.~Ahmad {\it et al.}  [SNO Collaboration],
  Phys.\ Rev.\ Lett.\  {\bf 89}, 011301 (2002)
  [arXiv:nucl-ex/0204008];
  S.~N.~Ahmed {\it et al.}  [SNO Collaboration],
  Phys.\ Rev.\ Lett.\  {\bf 92}, 181301 (2004)
  [arXiv:nucl-ex/0309004];
  K.~Eguchi {\it et al.}  [KamLAND Collaboration],
  Phys.\ Rev.\ Lett.\  {\bf 90}, 021802 (2003)
  [arXiv:hep-ex/0212021],
  P.~Adamson {\it et al.}  [MINOS Collaboration],
  Phys.\ Rev.\ Lett.\  {\bf 103}, 261802 (2009)
  [arXiv:0909.4996 [hep-ex]],
  arXiv:1007.2791 [hep-ex],
  A.~A.~Aguilar-Arevalo {\it et al.}  [MiniBooNE Collaboration],
  Phys.\ Rev.\ Lett.\  {\bf 103}, 061802 (2009)
  [arXiv:0903.2465 [hep-ex]].

\bibitem{DelNobile:2009st}
  E.~Del Nobile, R.~Franceschini, D.~Pappadopulo and A.~Strumia,
  Nucl.\ Phys.\  B {\bf 826}, 217 (2010)
  [arXiv:0908.1567 [hep-ph]].



\bibitem{multi_Higgs_models}
For examples:
  Y.~Grossman,
  Nucl.\ Phys.\  B {\bf 426}, 355 (1994)
  [arXiv:hep-ph/9401311] and references therein,
%
  Y.~Grossman and Y.~Nir,
  Phys.\ Lett.\  B {\bf 313}, 126 (1993)
  [arXiv:hep-ph/9306292],
%
  S.~Kanemura, T.~Kubota and E.~Takasugi,
  Phys.\ Lett.\  B {\bf 313}, 155 (1993)
  [arXiv:hep-ph/9303263],
%
  Y.~Okada, Y.~Shimizu and M.~Tanaka,
  Phys.\ Lett.\  B {\bf 405}, 297 (1997)
  [arXiv:hep-ph/9704223],
%
  A.~G.~Akeroyd, A.~Arhrib and E.~M.~Naimi,
  Phys.\ Lett.\  B {\bf 490}, 119 (2000)
  [arXiv:hep-ph/0006035],
%
  W.~Grimus and L.~Lavoura,
  Phys.\ Lett.\  B {\bf 546}, 86 (2002)
  [arXiv:hep-ph/0207229],
%
  S.~Kanemura, T.~Ota and K.~Tsumura,
  Phys.\ Rev.\  D {\bf 73}, 016006 (2006)
  [arXiv:hep-ph/0505191],
%
  E.~Lunghi and A.~Soni,
  JHEP {\bf 0709}, 053 (2007)
  [arXiv:0707.0212 [hep-ph]].
  
\bibitem{rad_seesaw_eg}
   M.~Aoki, S.~Kanemura and O.~Seto,
  Phys.\ Rev.\ Lett.\  {\bf 102}, 051805 (2009)
  [arXiv:0807.0361 [hep-ph]],
%
  M.~Aoki, S.~Kanemura and O.~Seto,
  Phys.\ Rev.\  D {\bf 80}, 033007 (2009)
  [arXiv:0904.3829 [hep-ph]],
%
  K.~L.~McDonald and B.~H.~J.~McKellar,
  arXiv:hep-ph/0309270.


\bibitem{dileptons1}
  F.~Cuypers and S.~Davidson,
  Eur.\ Phys.\ J.\  C {\bf 2}, 503 (1998)
  [arXiv:hep-ph/9609487],
%
  Y.~Kuno and Y.~Okada,
  Rev.\ Mod.\ Phys.\  {\bf 73}, 151 (2001)
  [arXiv:hep-ph/9909265] and references therein.

\bibitem{dileptons2}  
For examples (dilepton models):
%
  P.~H.~Frampton and D.~Ng,
  Phys.\ Rev.\  D {\bf 45}, 4240 (1992),
%
  P.~H.~Frampton,
  Phys.\ Rev.\ Lett.\  {\bf 69}, 2889 (1992),
%
  P.~H.~Frampton, D.~Ng, T.~W.~Kephart and T.~C.~Yuan,
  Phys.\ Lett.\  B {\bf 317}, 369 (1993)
  [arXiv:hep-ph/9210271],
%
  A.~G.~Akeroyd, M.~Aoki and H.~Sugiyama,
  Phys.\ Rev.\  D {\bf 79}, 113010 (2009)
  [arXiv:0904.3640 [hep-ph]].



\bibitem{babu_zee}
For examples (Zee/Babu or other radiative seesaw models):
  A.~Zee,
  Nucl.\ Phys.\  B {\bf 264}, 99 (1986),
%
  K.~S.~Babu,
  Phys.\ Lett.\  B {\bf 203}, 132 (1988),
%
  S.~Kanemura, T.~Kasai, G.~L.~Lin, Y.~Okada, J.~J.~Tseng and C.~P.~Yuan,
  Phys.\ Rev.\  D {\bf 64}, 053007 (2001)
  [arXiv:hep-ph/0011357],
%
  A.~Ghosal, Y.~Koide and H.~Fusaoka,
  Phys.\ Rev.\  D {\bf 64}, 053012 (2001)
  [arXiv:hep-ph/0104104],
%
  L.~M.~Krauss, S.~Nasri and M.~Trodden,
  Phys.\ Rev.\  D {\bf 67}, 085002 (2003)
  [arXiv:hep-ph/0210389].


\bibitem{doubly_Higg}
For examples (doubly charged Higgs):
%
  T.~G.~Rizzo,
  Phys.\ Rev.\  D {\bf 25}, 1355 (1982)
  [Addendum-ibid.\  D {\bf 27}, 657 (1983)],
%
  M.~L.~Swartz,
  Phys.\ Rev.\  D {\bf 40}, 1521 (1989),
%
  A.~G.~Akeroyd,
  Phys.\ Lett.\  B {\bf 353}, 519 (1995),
%
  E.~J.~Chun, K.~Y.~Lee and S.~C.~Park,
  Phys.\ Lett.\  B {\bf 566}, 142 (2003)
  [arXiv:hep-ph/0304069],
%
  A.~G.~Akeroyd and M.~Aoki,
  Phys.\ Rev.\  D {\bf 72}, 035011 (2005)
  [arXiv:hep-ph/0506176],
%
  C.~S.~Chen, C.~Q.~Geng and J.~N.~Ng,
  Phys.\ Rev.\  D {\bf 75}, 053004 (2007)
  [arXiv:hep-ph/0610118],
%
  C.~S.~Chen, C.~Q.~Geng and D.~V.~Zhuridov,
  Phys.\ Lett.\  B {\bf 666}, 340 (2008)
  [arXiv:0801.2011 [hep-ph]],
%
  P.~Fileviez Perez, T.~Han, G.~Y.~Huang, T.~Li and K.~Wang,
  Phys.\ Rev.\  D {\bf 78}, 071301 (2008)
  [arXiv:0803.3450 [hep-ph]],
%
  P.~Fileviez Perez, T.~Han, G.~Y.~Huang, T.~Li and K.~Wang,
  Phys.\ Rev.\  D {\bf 78}, 015018 (2008)
  [arXiv:0805.3536 [hep-ph]],
%
  A.~G.~Akeroyd and C.~W.~Chiang,
  Phys.\ Rev.\  D {\bf 80}, 113010 (2009)
  [arXiv:0909.4419 [hep-ph]].




\bibitem{type2seesaw}
For examples (type II seesaw):
%
  G.~Senjanovic and R.~N.~Mohapatra,
  Phys.\ Rev.\  D {\bf 12}, 1502 (1975),
%
  W.~Konetschny and W.~Kummer,
  Phys.\ Lett.\  B {\bf 70}, 433 (1977),
%
  T.~P.~Cheng and L.~F.~Li,
  Phys.\ Rev.\  D {\bf 22}, 2860 (1980),
%
  M.~Magg and C.~Wetterich,
  Phys.\ Lett.\  B {\bf 94}, 61 (1980),
%
  J.~Schechter and J.~W.~F.~Valle,
  Phys.\ Rev.\  D {\bf 22}, 2227 (1980),
%
  G.~Lazarides, Q.~Shafi and C.~Wetterich,
  Nucl.\ Phys.\  B {\bf 181}, 287 (1981),
%
  C.~Wetterich,
  Nucl.\ Phys.\  B {\bf 187}, 343 (1981),
%
  R.~N.~Mohapatra and G.~Senjanovic,
  Phys.\ Rev.\  D {\bf 23}, 165 (1981).


\bibitem{Abada:2007ux}
  A.~Abada, C.~Biggio, F.~Bonnet, M.~B.~Gavela and T.~Hambye,
  JHEP {\bf 0712}, 061 (2007)
  [arXiv:0707.4058 [hep-ph]].



\bibitem{babu_zee2}
  C.~S.~Chen, C.~Q.~Geng, J.~N.~Ng and J.~M.~S.~Wu,
  JHEP {\bf 0708}, 022 (2007)
  [arXiv:0706.1964 [hep-ph]],
%
  C.~S.~Chen, C.~Q.~Geng and D.~V.~Zhuridov,
  Eur.\ Phys.\ J.\  C {\bf 60}, 119 (2009)
  [arXiv:0803.1556 [hep-ph]],
%
  T.~Ohlsson, T.~Schwetz and H.~Zhang,
  Phys.\ Lett.\  B {\bf 681}, 269 (2009)
  [arXiv:0909.0455 [hep-ph]].

\bibitem{type1seesaw}
  P.~Minkowski,
  Phys.\ Lett.\ B {\bf 67}, 421 (1977),
%
  T.~Yanagida, in {\it Workshop on Unified Theories}, KEK report 79-18 p.95 (1979),
%
  M.~Gell-Mann, P.~Ramond, R.~Slansky, 
  in {\it Supergravity} (North Holland, Amsterdam, 1979)
  eds. P.~van~Nieuwenhuizen, D.~Freedman, p.315.,
%
  S.~L.~Glashow, in {\it 1979 Cargese Summer Institute on Quarks and Leptons} (Plenum Press,
  New York, 1980) eds. M.~Levy, J.-L.~Basdevant, D.~Speiser, J.~Weyers, R.~Gastmans and 
  M.~Jacobs, p.687;
%
  R.~Barbieri, D.~V.~Nanopoulos, G.~Morchio and F.~Strocchi,
  Phys.\ Lett.\ B {\bf 90}, 91 (1980),
%
  R.~N.~Mohapatra and G.~Senjanovic,
  Phys.\ Rev.\ Lett.\  {\bf 44}, 912 (1980).

\bibitem{type1set2}
  F.~del Aguila, J.~A.~Aguilar-Saavedra and R.~Pittau,
  JHEP {\bf 0710}, 047 (2007)
  [arXiv:hep-ph/0703261],
%
  A.~Atre, T.~Han, S.~Pascoli and B.~Zhang,
  JHEP {\bf 0905}, 030 (2009)
  [arXiv:0901.3589 [hep-ph]] and references therein,
%
  S.~Kovalenko, Z.~Lu and I.~Schmidt,
  Phys.\ Rev.\  D {\bf 80}, 073014 (2009)
  [arXiv:0907.2533 [hep-ph]].

\bibitem{Biggio:2008in}
  C.~Biggio,
  Phys.\ Lett.\  B {\bf 668}, 378 (2008)
  [arXiv:0806.2558 [hep-ph]],
%
  W.~Chao,
  arXiv:0806.0889 [hep-ph].
  
\bibitem{type1_3}
  F.~del Aguila and J.~A.~Aguilar-Saavedra,
  Phys.\ Lett.\  B {\bf 672}, 158 (2009)
  [arXiv:0809.2096 [hep-ph]].
  


\bibitem{type3seesaw}
  R.~Foot, H.~Lew, X.~G.~He and G.~C.~Joshi,
  Z.\ Phys.\  C {\bf 44}, 441 (1989),
%
  E.~Ma,
  Phys.\ Rev.\ Lett.\  {\bf 81}, 1171 (1998)
  [arXiv:hep-ph/9805219],
%
  E.~Ma and D.~P.~Roy,
  Nucl.\ Phys.\  B {\bf 644}, 290 (2002)
  [arXiv:hep-ph/0206150],
%
  B.~Bajc and G.~Senjanovic,
  JHEP {\bf 0708}, 014 (2007)
  [arXiv:hep-ph/0612029],
%
  B.~Bajc, M.~Nemevsek and G.~Senjanovic,
  Phys.\ Rev.\  D {\bf 76}, 055011 (2007)
  [arXiv:hep-ph/0703080],
%
  X.~G.~He and S.~Oh,
  JHEP {\bf 0909}, 027 (2009)
  [arXiv:0902.4082 [hep-ph]] and references therein,
%
  A.~Arhrib, R.~Benbrik and C.~H.~Chen,
  Phys.\ Rev.\  D {\bf 81}, 113003 (2010)
  [arXiv:0903.1553 [hep-ph]],
%



\bibitem{Abada:2008ea}
  A.~Abada, C.~Biggio, F.~Bonnet, M.~B.~Gavela and T.~Hambye,
  Phys.\ Rev.\  D {\bf 78}, 033007 (2008)
  [arXiv:0803.0481 [hep-ph]].


\bibitem{4thgen}
Some examples:
  I.~F.~Ginzburg, I.~P.~Ivanov and A.~Schiller,
  Phys.\ Rev.\  D {\bf 60}, 095001 (1999)
  [arXiv:hep-ph/9802364],
%
  P.~H.~Frampton, P.~Q.~Hung and M.~Sher,
  Phys.\ Rept.\  {\bf 330}, 263 (2000)
  [arXiv:hep-ph/9903387],
%
  M.~Maltoni, V.~A.~Novikov, L.~B.~Okun, A.~N.~Rozanov and M.~I.~Vysotsky,
  Phys.\ Lett.\  B {\bf 476}, 107 (2000)
  [arXiv:hep-ph/9911535] and references therein,
%
  V.~A.~Novikov, L.~B.~Okun, A.~N.~Rozanov and M.~I.~Vysotsky,
  Phys.\ Lett.\  B {\bf 529}, 111 (2002)
  [arXiv:hep-ph/0111028],
%
  T.~Cuhadar-Donszelmann, M.~Karagoz, V.~E.~Ozcan, S.~Sultansoy and G.~Unel,
  JHEP {\bf 0810}, 074 (2008)
  [arXiv:0806.4003 [hep-ph]],
%
  O.~Antipin, M.~Heikinheimo and K.~Tuominen,
  JHEP {\bf 0910}, 018 (2009)
  [arXiv:0905.0622 [hep-ph]].


\bibitem{dilepton_boson}
%
  P.~H.~Frampton and B.~H.~Lee,
  Phys.\ Rev.\ Lett.\  {\bf 64}, 619 (1990),
%
  P.~H.~Frampton and T.~W.~Kephart,
  Phys.\ Rev.\  D {\bf 42}, 3892 (1990),
%
  H.~Fujii, S.~Nakamua and K.~Sasaki,
  Phys.\ Lett.\  B {\bf 299}, 342 (1993),
%
  K.~Sasaki,
  Phys.\ Lett.\  B {\bf 308}, 297 (1993),
%
  P.~Ko, in
 {\it Constraint on dilepton gauge bosons from muonium $\to$ anti-muonium
  conversion}, UMN-TH-1134-93 \& TPI-MINN-93-26-T, Jun 7, 1993,
%
  M.~B.~Tully and G.~C.~Joshi,
  Phys.\ Lett.\  B {\bf 466}, 333 (1993)
  [arXiv:hep-ph/9905552],
%
  B.~Dutta and S.~Nandi,
  Phys.\ Lett.\  B {\bf 340}, 86 (1994),
%
  K.~Sasaki, K.~Tokushuku, S.~Yamada and Y.~Yamazaki,
  Phys.\ Lett.\  B {\bf 345}, 495 (1995),
%
  K.~Horikawa and K.~Sasaki,
  Phys.\ Rev.\  D {\bf 53}, 560 (1996)
  [arXiv:hep-ph/9504218],
%
  F.~Cuypers and M.~Raidal,
  Nucl.\ Phys.\  B {\bf 501}, 3 (1997)
  [arXiv:hep-ph/9704224],
%
  P.~H.~Frampton and M.~Harada,
  Phys.\ Rev.\  D {\bf 58}, 095013 (1998)
  [arXiv:hep-ph/9711448],
%
  N.~A.~Ky, H.~N.~Long and D.~V.~Soa,
  Phys.\ Lett.\  B {\bf 486}, 140 (2000)
  [arXiv:hep-ph/0007010],
%
  G.~Tavares-Velasco and J.~J.~Toscano,
  Phys.\ Rev.\  D {\bf 65}, 013005 (2002)
  [arXiv:hep-ph/0108114],
%
  E.~Ramirez Barreto, Y.~A.~Coutinho and J.~Sa Borges,
  Braz.\ J.\ Phys.\  {\bf 38}, 495 (2008),
%
  D.~Van Soa, P.~Van Dong, T.~T.~Huong and H.~N.~Long,
  J.\ Exp.\ Theor.\ Phys.\  {\bf 108}, 757 (2009)
  [arXiv:0805.4456 [hep-ph]].



\bibitem{Babu:2009aq}
  K.~S.~Babu, S.~Nandi and Z.~Tavartkiladze,
  Phys.\ Rev.\  D {\bf 80}, 071702 (2009)
  [arXiv:0905.2710 [hep-ph]].


\bibitem{Antusch:2006vwa}
  S.~Antusch, C.~Biggio, E.~Fernandez-Martinez, M.~B.~Gavela and J.~Lopez-Pavon,
  JHEP {\bf 0610}, 084 (2006)
  [arXiv:hep-ph/0607020].


\bibitem{Nakamura:2010zzi}
  K.~Nakamura  [Particle Data Group],
  J.\ Phys.\ G {\bf 37}, 075021 (2010).

\bibitem{Bernabeu:1993ta}
  J.~Bernabeu, E.~Nardi and D.~Tommasini,
  Nucl.\ Phys.\  B {\bf 409}, 69 (1993)
  [arXiv:hep-ph/9306251].


\bibitem{Marciano:2008zz} 
  W.~J.~Marciano, T.~Mori and J.~M.~Roney,
  Ann.\ Rev.\ Nucl.\ Part.\ Sci.\  {\bf 58}, 315 (2008) and references therein.

\bibitem{exp_MEG}
  S.~Ritt  [MEG Collaboration],
  Nucl.\ Phys.\ Proc.\ Suppl.\  {\bf 162}, 279 (2006),
%
  J.~Adam {\it et al.}  [MEG collaboration],
  Nucl.\ Phys.\  B {\bf 834}, 1 (2010)
  [arXiv:0908.2594 [hep-ex]].
%
\bibitem{Brooks:1999pu}
  M.~L.~Brooks {\it et al.}  [MEGA Collaboration],
  Phys.\ Rev.\ Lett.\  {\bf 83}, 1521 (1999)
  [arXiv:hep-ex/9905013].



\bibitem{exp_SuperB} 
  T.~Aushev {\it et al.},
  KEK-REPORT-2009-12, Feb 2010,
  [arXiv:1002.5012 [hep-ex]] and references therein.


\bibitem{Cheng:1985bj}
  T.~P.~Cheng and L.~F.~Li,
  {\it Gauge Theory Of Elementary Particle Physics},
 Oxford, UK: Clarendon (1984) 536p. (Oxford Science Publications).


\bibitem{mu2eg_SM}
  T.~P.~Cheng and L.~F.~Li,
  Phys.\ Rev.\  D {\bf 16}, 1425 (1977),
%
  S.~T.~Petcov,
  Sov.\ J.\ Nucl.\ Phys.\  {\bf 25}, 340 (1977)
  [Yad.\ Fiz.\  {\bf 25}, 641 (1977)]
  [Erratum-ibid.\  {\bf 25}, 698 (1977)]
  [Erratum-ibid.\  {\bf 25}, 1336 (1977)],
%
  W.~J.~Marciano and A.~I.~Sanda,
  Phys.\ Lett.\  B {\bf 67}, 303 (1977),
%
  B.~W.~Lee and R.~E.~Shrock,
  Phys.\ Rev.\  D {\bf 16}, 1444 (1977).


\bibitem{Fujikawa:1972fe}
  K.~Fujikawa, B.~W.~Lee and A.~I.~Sanda,
  Phys.\ Rev.\  D {\bf 6}, 2923 (1972).

\bibitem{Peskin:1995ev}
  M.~E.~Peskin and D.~V.~Schroeder,
  {\it An Introduction To Quantum Field Theory},
 Reading, USA: Addison-Wesley (1995) 842p.


\bibitem{Bilenky:1987ty}
  S.~M.~Bilenky and S.~T.~Petcov,
  Rev.\ Mod.\ Phys.\  {\bf 59}, 671 (1987)
  [Erratum-ibid.\  {\bf 61}, 169 (1989)]
  [Erratum-ibid.\  {\bf 60}, 575 (1988)].


\bibitem{Barate:2003sz}
  R.~Barate {\it et al.}  [LEP Working Group for Higgs boson searches and
                  ALEPH Collaboration and  and],
  Phys.\ Lett.\  B {\bf 565}, 61 (2003)
  [arXiv:hep-ex/0306033].


\bibitem{11of_ref22}
  I.~Sick and J.~S.~Mccarthy,
  Nucl.\ Phys.\  A {\bf 150}, 631 (1970),
%
  T.~W.~Donnelly and J.~D.~Walecka,
  Ann.\ Rev.\ Nucl.\ Part.\ Sci.\  {\bf 25}, 329 (1975),
%
  B.~Frois and C.~N.~Papanicolas,
  Ann.\ Rev.\ Nucl.\ Part.\ Sci.\  {\bf 37}, 133 (1987)
and references therein.


\bibitem{Dohmen:1993mp}
  C.~Dohmen {\it et al.}  [SINDRUM II Collaboration.],
  Phys.\ Lett.\  B {\bf 317}, 631 (1993).

\bibitem{Bertl:2006up}
  W.~H.~Bertl {\it et al.}  [SINDRUM II Collaboration],
  Eur.\ Phys.\ J.\  C {\bf 47}, 337 (2006).

\bibitem{Kitano:2002mt}
  R.~Kitano, M.~Koike and Y.~Okada,
  Phys.\ Rev.\  D {\bf 66}, 096002 (2002)
  [Erratum-ibid.\  D {\bf 76}, 059902 (2007)]
  [arXiv:hep-ph/0203110].


\bibitem{Zeff}
  J.~C.~Sens,
  Phys.\ Rev.\  {\bf 113}, 679 (1959),
%
  H.~C.~Chiang, E.~Oset, T.~S.~Kosmas, A.~Faessler and J.~D.~Vergados,
  Nucl.\ Phys.\  A {\bf 559}, 526 (1993).

\bibitem{Suzuki:1987jf}
  T.~Suzuki, D.~F.~Measday and J.~P.~Roalsvig,
  Phys.\ Rev.\  C {\bf 35}, 2212 (1987).




\bibitem{exp_COMET}
  A.~Sato,
  PoS {\bf NUFACT08}, 105 (2008),
%
  and talk given at the 4th International Workshop on Nuclear and Particle Physics at J-PARC (NP08),
  Mito, Ibaraki, Japan, March 2008 (http://nuclpart.kek.jp/NP08/presentations/ muon/pdf/NP08\_Muon\_Sato.pdf),
  Y.~G.~Cui {\it et al.}  [COMET Collaboration],
  KEK-2009-10, Jun 2009.
  
\bibitem{exp_mu2e}
  R.~M.~Carey {\it et al.}  [Mu2e Collaboration],
  FERMILAB-PROPOSAL-0973, Oct 2008,
%
  D.~Glenzinski,
  AIP Conf.\ Proc.\  {\bf 1222}, 383 (2010).

\bibitem{g-2_review1}
See for examples:
%
  M.~Davier and W.~J.~Marciano,
  Ann.\ Rev.\ Nucl.\ Part.\ Sci.\  {\bf 54}, 115 (2004),
%
  M.~Passera,
  J.\ Phys.\ G {\bf 31}, R75 (2005)
  [arXiv:hep-ph/0411168],
%
  M.~Passera,
  Nucl.\ Phys.\ Proc.\ Suppl.\  {\bf 169}, 213 (2007)
  [arXiv:hep-ph/0702027],
%
  J.~P.~Miller, E.~de Rafael and B.~L.~Roberts,
  Rept.\ Prog.\ Phys.\  {\bf 70}, 795 (2007)
  [arXiv:hep-ph/0703049],
%
  F.~Jegerlehner and A.~Nyffeler,
  Phys.\ Rept.\  {\bf 477}, 1 (2009)
  [arXiv:0902.3360 [hep-ph]].
%

\bibitem{mg-2_SM}
M.~Passera, W.~J.~Marciano and A.~Sirlin,
  arXiv:1001.4528 [hep-ph].


\bibitem{mg-2_exp1}
  G.~W.~Bennett {\it et al.}  [Muon G-2 Collaboration],
  Phys.\ Rev.\  D {\bf 73}, 072003 (2006)
  [arXiv:hep-ex/0602035],
%
  Phys.\ Rev.\ Lett.\  {\bf 92}, 161802 (2004)
  [arXiv:hep-ex/0401008],
%
  Phys.\ Rev.\ Lett.\  {\bf 89}, 101804 (2002)
  [Erratum-ibid.\  {\bf 89}, 129903 (2002)]
  [arXiv:hep-ex/0208001].


\bibitem{mg-2_exp2}
  B.~L.~Roberts,
  arXiv:1001.2898 [hep-ex].

\bibitem{eg-2_exp}
  D.~Hanneke, S.~Fogwell and G.~Gabrielse,
  Phys.\ Rev.\ Lett.\  {\bf 100}, 120801 (2008)
  [arXiv:0801.1134 [physics.atom-ph]].


\bibitem{eg-2_fineS}
  R.~S.~Van Dyck, P.~B.~Schwinberg and H.~G.~Dehmelt,
  Phys.\ Rev.\ Lett.\  {\bf 59}, 26 (1987),
%
  G.~Gabrielse, D.~Hanneke, T.~Kinoshita, M.~Nio and B.~C.~Odom,
  Phys.\ Rev.\ Lett.\  {\bf 97}, 030802 (2006)
  [Erratum-ibid.\  {\bf 99}, 039902 (2007)].


\bibitem{DELPHI}
  J.~Abdallah {\it et al.}  [DELPHI Collaboration],
  Eur.\ Phys.\ J.\  C {\bf 35}, 159 (2004)
  [arXiv:hep-ex/0406010].



\end{thebibliography}
\end{document}